\documentclass[paper]{JFM-FLM_Au}
\usepackage{graphicx}
\usepackage{rotating}
\usepackage{amsmath}
\usepackage{amssymb}
\usepackage{hyperref}
\usepackage[usenames,dvipsnames,svgnames,table]{xcolor}
\usepackage{mathptmx} % comment out on mac
\usepackage[utf8]{inputenc}
\usepackage{newtxmath}
\usepackage{pgfplots}
\usepackage{natbib}
\usepackage{microtype}
\usepackage{diagbox}

\hypersetup{
colorlinks=true,
linkcolor=blue,
citecolor=blue
}
\makeatletter

\patchcmd{\NAT@citex}
  {\@citea\NAT@hyper@{%
     \NAT@nmfmt{\NAT@nm}%
     \hyper@natlinkbreak{\NAT@aysep\NAT@spacechar}{\@citeb\@extra@b@citeb}%
     \NAT@date}}
  {\@citea\NAT@nmfmt{\NAT@nm}%
   \NAT@aysep\NAT@spacechar\NAT@hyper@{\NAT@date}}{}{}

\patchcmd{\NAT@citex}
  {\@citea\NAT@hyper@{%
     \NAT@nmfmt{\NAT@nm}%
     \hyper@natlinkbreak{\NAT@spacechar\NAT@@open\if*#1*\else#1\NAT@spacechar\fi}%
       {\@citeb\@extra@b@citeb}%
     \NAT@date}}
  {\@citea\NAT@nmfmt{\NAT@nm}%
   \NAT@spacechar\NAT@@open\if*#1*\else#1\NAT@spacechar\fi\NAT@hyper@{\NAT@date}}
  {}{}

\makeatother

\newcommand{\vect}[1]{\boldsymbol{#1}}

\newcommand\Fr{\mbox{\textit{Fr}}} % Froude number
\newcommand\im{\mathrm{i}\mkern1mu} % imaginary unit
\newcommand\diff{\mathrm{d}} % differential
\newcommand\shape{\chi} % shape factor
\newcommand\shapefu{\alpha} % shape factor coeff of du/dx
\newcommand\shapefs{\beta} % shape factor coeff of ds/dx
\DeclareMathOperator\real{Re}
\DeclareMathOperator\imag{Im}

\title{Channel confinement mollifies roll wave instabilities}
\lefttitle{J.\ Langham, C.\ Gadal, J.P.\ Webb, C.G.\ Johnson, J.M.N.T.\ Gray}
\author{Jake Langham\aff{1}, Cyril Gadal\aff{1}, Jamie P.\ Webb\aff{1}, Chris G.\ Johnson\aff{1} \and J.M.N.T.\ Gray\aff{1}}
\affiliation{\aff{1}Department of Mathematics and Manchester Centre for
Nonlinear Dynamics,
University of Manchester,
Manchester, M13 9PL, UK
}

\corresau{Jake Langham, \email{jacob.langham@manchester.ac.uk}}

\begin{document}
%\linenumbers
\maketitle

\begin{abstract}
Under certain conditions, gravity currents spontaneously develop large-amplitude
streamwise undulations on their free surfaces (‘roll waves’).  The effect of
channel geometry on this instability is investigated herein. Via a
section-averaged stability analysis conducted independently of the rheology of
the flowing material, convex open channels are found to be stabilising, relative
to unconfined flows. This is shown to agree with experimental observations of
laterally shallow currents of water and dry granular material in trapezoidal
channels. For such flows, the influence of the geometry can be characterised by
a single dimensionless parameter, which predicts stabilisation as the channel
width narrows. Including the cross-stream velocity profile in the analysis
further stabilises predictions and can render steady flows in triangular
channels unconditionally stable – a finding borne out by our experiments.
Consequently, laterally tilting a trapezoidal channel can also stabilise flows
by adjusting the wetted region towards a triangular one. Implications for
observations in natural channels are considered by adapting an existing model of
roll waves in the Illgraben, Switzerland, to incorporate channel geometry. The
resulting simulations produce suitable waves for a realistic cross-section, but
are stabilised by a modest narrowing of the channel. Complementary nonlinear
wave solutions are then constructed and used to show that channel confinement
also diminishes amplitudes of observed waves. Finally, exceptions to the
analysis are discussed, including the possibility for static material to shield
waves from the effects of geometry. We demonstrate this using observations of
undamped granular avalanches travelling through an arrested deposit in a
triangular chute.
\end{abstract}

\maketitle

\section{Introduction}
\label{sec:intro}
It has long been recognised that steady inclined free-surface flows can be
vulnerable to a linear instability that promotes the spontaneous development of
streamwise-travelling surges.
These are commonly called `roll waves' in the fluid
mechanics literature and were originally documented in turbulent watercourses by~\cite{Cornish1934}. Numerous theoretical and laboratory studies have
since demonstrated that essentially the same instability underlies waves found
in laminar Newtonian flows~\citep{Benjamin1957,Yih1963,Yu2024}, as well as in dry
granular media~\citep{Forterre2003,Forterre2006,Viroulet2018} and various complex
fluids~\citep{Liu1994,Ng1994,Balmforth2004a,RuyerQuil2012,Allouche2017}.
Meanwhile, increasingly sophisticated monitoring efforts by field geologists
have documented the prevalence of roll waves in the environment~\citep{Li2024},
most notably in alpine flows of debris, where they exacerbate the risks
associated with seasonal natural
hazards~\citep{Schoffl2023,Aaron2025,Spielmann2025}.
The basic ingredients of the instability are a shallow inertial flow of
approximately homogeneous fluid that is driven along a slope by gravity and
resisted by friction at the base of the flow.  Instability occurs when
power input to small-amplitude disturbances from gravitational forcing exceeds
that which can be dissipated by the basal stresses, causing them to
grow~\citep{Trowbridge1987}.  Beyond this, there are routes to
phenomenologically similar waves where the destabilising mechanism is either
somewhat more complicated, or qualitatively different, such as in the case of
very thin laminar films (where surface tension is
important)~\cite{Yih1963,Liu1993}, the inertialess instability of
shear-thickening fluids~\citep{DarboisTexier2020,Balmforth2025}, yield-stress
fluids featuring unyielded plugs~\citep{Liu1994,Balmforth2004a} and granular
avalanching (`erosion--deposition') waves triggered by finite mass
releases~\citep{Borzsonyi2008,Takagi2011,Edwards2015,Rocha2019}.  Though these
latter cases are also sometimes called roll waves, they will not be considered
explicitly in this paper.

Over the years, many studies have tackled the problem of identifying critical
conditions for particular fluids to feature a roll wave instability. For
simplicity, theoretical analyses overwhelmingly situate the flow on an idealised
constant incline and impose uniformity in the cross-slope
direction~\citep[e.g.][]{Jeffreys1925,Dressler1949,Benjamin1957,Yih1963,Forterre2003,Forterre2006,Trowbridge1987}.
To match these predictions, experiments are typically conducted in rectangular
channels, designed with a cross-section that is much wider than the flow
depth~\citep{Forterre2003,Liu2005,Allouche2017,Viroulet2018,Noma2021}. 
However, outside the laboratory,
flows are necessarily confined to natural or man-made channels that
often violate these assumptions. 
Consequently, a few authors have analysed the
stability of section-averaged flow equations, which embed the effects of channel
geometry on mass and momentum conservation into their
formulation~\citep{Craya1952,Iwasa1954,Ponce1995,Lamri2020}. These studies are
usually framed from the perspective of informing the design of conduits that are stable
to roll waves and focus their attention on obtaining accurate formulae for
currents of laminar and turbulent water.  Of particular interest in this setting
is the analysis of~\cite{Iwasa1954}, whose early derivations of stability
bounds for water flow in general channels include a passage on trapezoidal
cross-sections, 
which demonstrates that narrower trapezoids are less vulnerable to instability than
wider ones. The paper concludes with the following sentence:
%, over 70 years ago.
%
%Unfortunately, this work has not been widely cited outside of the hydraulic
%engineering literature.
\begin{quote}
    It is the author's hope that further experimental researches on this problem will
verify the above derived analytical result.
\end{quote}
To our knowledge, this appeal has not been addressed in the intervening 72
years.

\begin{figure}
    \includegraphics[width=\textwidth]{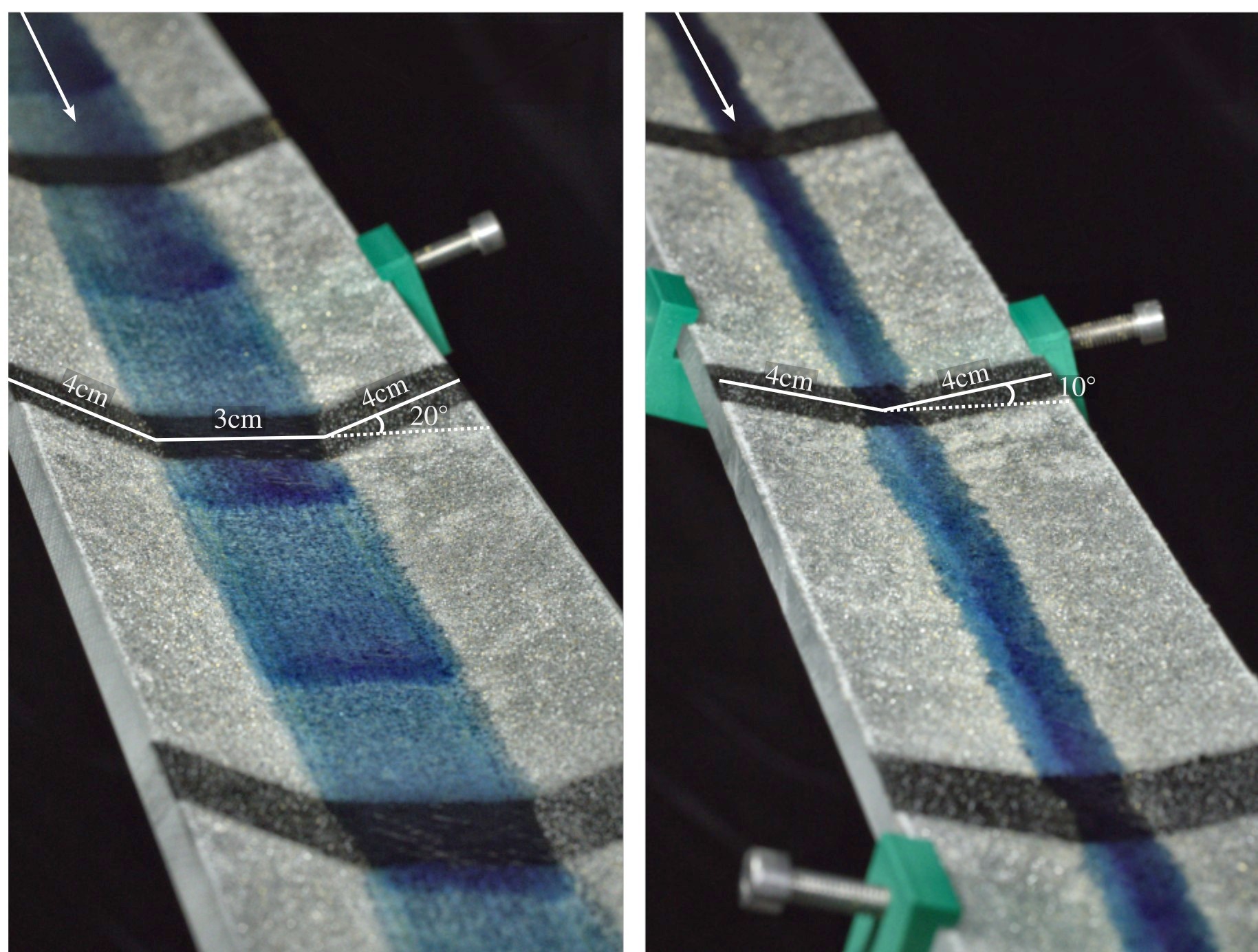}
    \caption{%
    Flows of water in trapezoidal (left) and triangular (right) channels, both
    inclined at $15^\circ$ to the horizontal. 
    Annotations depict the geometry of each channel cross-section perpendicular
    to the mean flow.
    Fluid is delivered via an
    upstream outlet (out of frame) at a constant mass flux of
    $10\mathrm{g\, /s}$ and thereafter driven by gravity (arrows point
    downstream). 
    A small quantity of blue dye has been added to
    aid visualisation. 
    Flow in the trapezium is spontaneously unstable -- several roll wave crests are
    visible in the image as localised concentrations of dye. 
    Conversely, flow in the triangular channel remains steady.
    }%
    \label{fig:introfig}%
\end{figure}

However, experimentation in our laboratory vindicates Iwasa's prediction.
Figure~\ref{fig:introfig} shows an example photograph of water flowing in a $3\mathrm{m}$-long
inclined channel with trapezoidal cross-section,
%that is much wider than the
%mean flow depth, 
alongside flow at the same mean flux in a triangular channel.
While the trapezium flow readily produced prominent roll waves, there were no
visible surface waves in the triangular channel.
Subsequent attempts to generate roll waves in the latter case by
adjusting the flux and slope angle all led to steady flows. Furthermore, manually
imparted disturbances decayed back to the base flow.
This suggests that channel geometry places a strong control on wave development.

Given the burgeoning interest in roll waves in natural flows whose rheologies
may be complex or unknown \emph{a priori}, we revisit this problem, shifting our
emphasis away from the hydraulic engineering applications that
motivated prior studies.
General formulae are derived for the linear growth rates of flow disturbances
and corresponding neutral stability curves that may be adapted for single-phase
fluids with any basal resistance function.  We also show how to compute
nonlinear roll wave solutions in this framework, since this is relevant for
constraining the amplitudes and lengths of waves that may be observed under a
given set of flow conditions.  These analyses are illustrated for flows
that are laterally shallow (in the sense that their depth is appreciably smaller
than their cross-stream width) and situated in channels with trapezoidal and
power-law cross-sections, as well as horizontally tilted trapezoids, which are
relatively stabilising.

In the course of our investigation, we extend Iwasa's claim regarding the
stabilising effects of narrower channels to any fluid whose velocity in steady
state can be modelled by a power law, which includes dry granular flows and
simple descriptions of non-Newtonian fluids.  To support our theoretical
analysis, numerical simulations and experiments of dry granular flow are
conducted, which both demonstrate the stabilising role of channel confinement.
Most strikingly, just like the water in the figure~\ref{fig:introfig}
experiment, the granular flows in a triangular channel appear to be
unconditionally stable.  This is predicted by the linear theory, provided that
lateral variations in the flow velocity profile are properly accounted for.

\section{Linear analysis}
\label{sec:linear analysis}%
\begin{figure}
    \begin{center}
    \includegraphics[width=10cm]{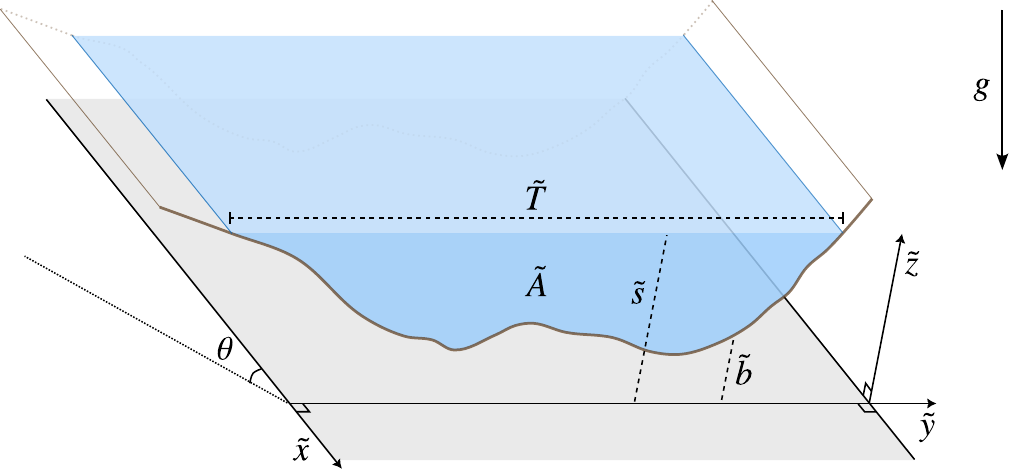}%
    \end{center}
    \caption{Sketch of the theoretical description,
    with some of the key geometric quantities labelled.
    }
    \label{fig:flow diagram}%
\end{figure}

We consider a fluid flowing within a straight open channel of invariant cross-section
that is inclined at an angle $\theta$ to the horizontal. The upper surface of the
flow is assumed to be stress free.
A diagram of the flow is
given in figure~\ref{fig:flow diagram} for reference.
We adopt a Cartesian coordinate frame aligned with the slope, with $\tilde x$
pointing downstream, $\tilde y$ across it and denote the channel base as
$\tilde b\equiv \tilde b(\tilde y)$.  
Later on, $\tilde{z}$ will be used for the upward slope-normal direction.
The area of the flow each time $\tilde t$ and downslope location
$\tilde x$ is defined by 
\begin{equation}
    \tilde A = \int_{-\infty}^{\infty} \tilde h\,\diff\tilde y,
    \label{eq:area}%
\end{equation}
where the field $\tilde h \equiv \max\{\tilde s - \tilde b, 0\}$ defines the
local flow depth, with $\tilde{s}\equiv\tilde{s}(\tilde{x},\tilde{t})$ denoting
the mean vertical elevation of the free surface measured above the same
reference plane that sets the bed height (as in figure~\ref{fig:flow diagram}).
%Without loss of generality, we set $\tilde{b}=0$ at the lowest point in the
%channel, so that $\max\tilde{h} = \tilde{s}$.
%
A natural characteristic length scale for the flow is
the top-width-averaged flow depth $\tilde D$,
known as the \emph{hydraulic depth} in engineering, given by
\begin{equation}
    \tilde D = 
    \frac{\tilde{A}}{\tilde{T}},
    %\frac{1}{\tilde{y}_2 - \tilde{y}_1}
    %\int_{-\infty}^{\infty} \tilde h \,\diff \tilde y,
%    =
%    \frac{\tilde A}{\tilde A'},
    \label{eq:hydraulic depth}%
\end{equation}
where $\tilde{T}\equiv\tilde{T}(\tilde{x},\tilde{t})$ is the top width of the
wetted region.
%Note that $\tilde{T} = \diff \tilde{A}/\diff \tilde{s}$. This will be
%used to simplify some expressions later.
%where $\tilde{y}_1$ and $\tilde{y}_2$ denote the lateral bounding points of the
%free surface in the $\tilde y$ direction and 

If the flow extends much further along the downslope direction
than across its depth or width, the equations
of motion for the flowing material may be reduced by
averaging them over the channel cross-section, since the
leading-order dynamics of mass and momentum occur over the longest length
scale. %~\citep{Chow1959,Henderson1966}. 
This leads to governing equations that capture the bulk evolution of mass and momentum
over the cross-section, posed in terms of
%mass and momentum posed in terms of
%at time $\tilde t$ and downslope location $\tilde x$, 
%the area
%occupied by flowing material 
$\tilde A(\tilde x,\tilde t)$ 
and the section-averaged downslope
velocity $\tilde U(\tilde x,\tilde t)$, defined as
\begin{equation}
    \tilde U = \frac{1}{\tilde A}\int_{\tilde \Omega} \tilde u
    \,\diff \tilde \Omega,
\end{equation}
where $\tilde u$ is the downslope flow velocity field and $\tilde \Omega\equiv \tilde
\Omega(\tilde x, \tilde t)$ denotes the region occupied by the fluid, with $|\tilde
\Omega|= \tilde A$.
The resulting system~\citep[see e.g.][]{Chow1959,Henderson1966} is:%, as derived in Appendix~\ref{appendix:derivation}:
\begin{subequations}
\begin{gather}
    \frac{\partial \tilde A}{\partial \tilde t} + \frac{\partial ~}{\partial
    \tilde x}
    \left(
    \tilde A \tilde U
    \right) = 0,\label{eq:governing A 1}\\
    \frac{\partial ~}{\partial \tilde t}(\tilde A \tilde U)
    + \frac{\partial ~}{\partial \tilde x}(\shape \tilde A \tilde U^2)
    + g \tilde A \cos(\theta) \frac{\partial \tilde s}{\partial \tilde x}
    = g \tilde A \sin(\theta) - \frac{\tilde\tau}{\tilde \rho},
    \label{eq:governing A 2}%
\end{gather}
    \label{eq:governing all}%
\end{subequations}
where $g$ is gravitational acceleration, 
%$\tilde s(\tilde x, \tilde t)$ is the
%mean vertical elevation of the free surface above some reference plane, 
$\tilde
\rho$ is the density of the flowing material (assumed constant herein), $\tilde
\tau$ is the total downslope drag exerted on the flow integrated over the
wetted perimeter and $\shape$ is a `shape factor' that arises when averaging over
the quadratic momentum term. This latter variable accounts for vertical and
cross-stream shear in the velocity profile and may be written as
\begin{equation}
    \shape 
    = \frac{1}{\tilde A \tilde U^2}\int_{\tilde \Omega} \tilde u^2 \,\diff \tilde\Omega
    =
    1 + \frac{1}{\tilde A} \int_{\tilde \Omega} \left(
    \frac{\tilde u}{\tilde U} - 1
    \right)^{\!2} \diff \tilde \Omega.
    \label{eq:shape factor}%
\end{equation}
Note that $\shape \geq 1$, with $\shape = 1$ if and only if $\tilde u = \tilde U$
everywhere (plug flow),
which is often assumed in applications 
\citep[see e.g.][]{Iverson1997}.
In confined flows, the velocity shear can often be significant. Therefore, we
retain $\shape$ in our analysis.
The drag function may be related to the basal shear
stress $\tilde \tau_b$ acting on the flow at each point along the wetted portion
of the channel via the line integral
\begin{equation}
    \tilde \tau = \int_{\mathcal{P}} \tilde \tau_b %(\tilde l)\,\diff \tilde l,
    \label{eq:tau integral}%
\end{equation}
where $\mathcal{P}$ denotes the curve bounding the underside of the flow.
Both $\shape$ and $\tilde{\tau}$ are treated as arbitrary functions of the
flow variables in this section.
%where $\gamma \equiv \gamma(\tilde x, \tilde t)$ is the curve along the
%cross-slope direction that delineates the portion of the channel in contact with
%the flowing material, parametrised by some curvilinear coordinate $\tilde l$.
%While $\tilde{\tau}_b$ depends on the specific local flow conditions
%at the base, it is typically modelled as a function of the mean velocity and
%flow depth.

As well as a shape factor and drag parametrisation, a function $\tilde{b}$
for the channel
shape must be given in order to close~\eqref{eq:governing A 1}
and~\eqref{eq:governing A 2}, thereby enabling the dependence of $\tilde A$ on
the free surface elevation to be determined. 
For now, we allow the shape to remain arbitrary, but assume that $\tilde{A}$ is continuously
differentiable injective function of $\tilde{s}$.
In the special case of a
rectangular channel of width $\tilde{w}$, $\tilde{A} = \tilde{h}\tilde{w}$,
%For any channel featuring a
%flat base of width $\tilde w$ at some $\tilde z = \tilde b_*$, then 
%$\tilde A/\tilde w \to \tilde s - \tilde b_*$ as $\tilde w\to \infty$,
and~\eqref{eq:governing A 1} and~\eqref{eq:governing A 2} reduce to
a one-dimensional depth-averaged shallow-layer model in $\tilde{h}$ and $\tilde{U}$.
Moreover, if $\tilde{w}\gg\tilde{h}$, then any drag from the channel sidewalls
may be justifiably neglected. These twin assumptions
are a common starting point for
the analysis of roll waves in the literature
\citep[e.g.][]{Jeffreys1925,Dressler1949,Trowbridge1987,Forterre2003}.
Here, we investigate the role that lateral confinement plays.

The simplest solutions to~\eqref{eq:governing A 1} and~\eqref{eq:governing
A 2} are steady uniform flows, where both $\tilde A = \tilde A_0$ and $\tilde U
= \tilde U_0$ are constant with respect to $\tilde x$ and $\tilde t$. 
The convention that `0' subscripts denote quantities evaluated at this state
is adopted throughout the paper, so $\tilde{s}_0$ signifies the steady
surface elevation, $\tilde{D}_0$ the steady hydraulic depth, and so on.
Steady uniform flows
occur when basal drag exactly resists the downslope gravitational forcing, i.e.\
\begin{equation}
    \tilde \rho g\tilde A_0\sin(\theta) = \tilde \tau(\tilde A_0, \tilde U_0) 
    \equiv \tilde \tau_0.
    \label{eq:steady balance}%
\end{equation}
This equilibrium is typically maintained in fluids via the adoption of vertical
and cross-slope shear profiles that transfer frictional resistance into material
stresses that counter the gravitational body force.
%We denote the corresponding free surface elevation for this state by $\tilde
%s_0$.  
%Throughout the paper, the free surface shall be measured from the lowest
%point in the channel, so that its value corresponds to the maximum depth of the
%flow.
%
%We denote the hydraulic depth in steady uniform conditions by $\tilde D_0$.

The scales given by the base flow may be used to bring the governing equations
into a convenient dimensionless form, by applying the transformations
\begin{subequations}
\begin{gather}
    t = \tilde t \tilde U_0/\tilde{\ell}, \quad
    x = \tilde x / \tilde{\ell}, \quad
    s = \tilde s / \tilde s_0, \quad
    D = \tilde D / \tilde D_0, \tag{\theequation\emph{a--d}}\\
    A = \tilde A / \tilde A_0, \quad
    U = \tilde U / \tilde U_0, \quad
    \tau = \tilde \tau / (\tilde \rho \tilde A_0 
    \tilde U_0^2/\tilde{\ell}),
    \tag{\theequation\emph{e--g}}%
\end{gather}
    \label{eq:nondimensionalisation}%
\end{subequations}
with $\tilde{\ell} \equiv \tilde U_0^2 / (g \sin\theta)$.  Since $\tilde{T} =
\diff \tilde A/\diff \tilde s$ by the geometry of the problem,
\eqref{eq:hydraulic depth} and~(\ref{eq:nondimensionalisation}\emph{c--e}) imply that $D =
(A\tilde s_0/\tilde D_0)/(\diff A/\diff s)$.
These observations, combined with the non-dimensionalisation are sufficient to
convert~\eqref{eq:governing A 1}
and~\eqref{eq:governing A 2}, into the following
semilinear system for $s$ and $U$:
\begin{subequations}
\begin{gather}
    \frac{\partial s}{\partial t}
    + U \frac{\partial s}{\partial x}
    + \frac{D}{\Delta} \frac{\partial U}{\partial x} = 0,\label{eq:governing nondim
    1}\\
    \frac{\partial U}{\partial t}
    + \left(U + \shapefu\right) \frac{\partial U}{\partial x}
    + \Delta \left(
        \frac{1}{\Fr^2} %+ \frac{(\shape - 1)U^2}{D}
        + \shapefs
        \right)\frac{\partial s}{\partial x} = 1 - \frac{\tau}{A}.
    \label{eq:governing nondim 2}%
\end{gather}
\end{subequations}
The above equations include two dimensionless control parameters -- the Froude
number of the base flow
$\Fr \equiv \tilde U_0/\sqrt{g\cos(\theta) \tilde D_0}$, and the length scale
ratio
$\Delta \equiv \tilde{s}_0/\tilde{D}_0$, %\end{equation}
while
\begin{subequations}
\begin{equation}
    \shapefu \equiv 2(\shape - 1)U + U^2 \frac{\partial \shape}{\partial U},
    \quad
    \shapefs \equiv \frac{(\shape - 1)U^2}{D} + \frac{U^2}{\Delta} \frac{\partial \shape}{\partial s}
    \tag{\theequation\emph{a,b}}%
\end{equation}
    \label{eq:alpha beta}%
\end{subequations}
are shape factor terms that vanish when $\shape \equiv 1$.  

Since the 
flow elevation is measured relative to an arbitrary reference height, 
the forthcoming analysis does not depend on the
absolute value of $\Delta$, which can be removed from the problem by
scaling $\tilde{s}$ by $\tilde{D}_0$ instead of $\tilde{s}_0$
in~(\ref{eq:nondimensionalisation}\emph{c}).
However, the choice made above draws out how the ensuing formulae vary as the
channel geometry changes, with the surface elevation held fixed.
The value of $\Delta$ describes how much the channel deviates
from a rectangular cross-section. 
We illustrate this in figure~\ref{fig:hydraulic
depth}.
\begin{figure}
    \includegraphics[width=\textwidth]{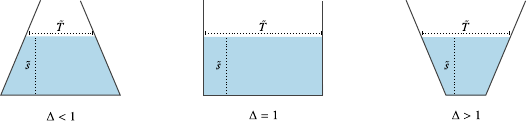}%
    \caption{Three channels with different values of $\Delta$.}
    \label{fig:hydraulic depth}%
\end{figure}
Without loss of generality, we always measure $\tilde{s}$ such that $\Delta = 1$
for a rectangular channel.  Typical convex open channels such as trapezoidal, or
semicircular cross sections then have $\Delta > 1$, since regardless of the
surface elevation, $\tilde{s}\tilde{T} > \tilde{A}$, implying
$\tilde{s}>\tilde{D}$.
Conversely, channels
with sidewalls that are banked with a negative gradient obey the opposite
inequalities, so $\Delta < 1$.
Depending on the specifics of the problem and particular modelling choices,
geometric parameters other than $\Delta$ may also be embedded within the
closures for $\shape$ and $\tau$.

At the steady uniform equilibrium~\eqref{eq:steady balance},
\begin{equation}
    s_0 = A_0 = D_0 = U_0 = \tau_0 = 1, \quad
    \frac{\mathrm{d}A}{\mathrm{d}s_0} = \Delta,
\end{equation}
where henceforth, for notational convenience, we write
$\frac{\mathrm{d}A}{\mathrm{d}s_0} =
\big(\frac{\mathrm{d}A}{\mathrm{d}s}\big)_0$ and so on, for derivatives
evaluated at the base state.
The linear stability properties of this steady flow may be determined by
assessing its response to a complete set of small disturbances. We write
\begin{gather}
    s = 1 + s_1 \exp(\im k x + \sigma t), \quad
    U = 1 + U_1 \exp(\im k x + \sigma t),
    \label{eq:modes}%
\end{gather}
where $|s_1|, |U_1| \ll 1$ are unknown complex amplitudes, $k$ is an arbitrary
real-valued wavenumber and $\sigma$ is a complex number whose real part
defines the temporal growth rate of a given mode.
The phase speed $c$ of the modes is $c = -\imag(\sigma)/k$.
Substituting these expressions into~\eqref{eq:governing
nondim 1} and~\eqref{eq:governing nondim 2}, dropping nonlinear terms and
rearranging yields the following formula for $\sigma$ as a function of $k$:
\begin{equation}
    \sigma = -\im k \left(1 + \frac{\shapefu_0}{2}\right)
    - \frac{1}{2}
        \frac{\partial \tau}{\partial U_0}
        \pm
        \sqrt{
            \left(\frac{1}{2}\frac{\partial \tau}{\partial U_0}\right)^{\!2}
            -k^2\left(
            \frac{1}{\Fr^2} + \frac{\shapefu_0^2}{4} + \shapefs_0
            \right)
            + \im k
            \left(
            \frac{1}{\Delta}\frac{\partial \tau}{\partial s_0}
            - 1
            + \frac{\shapefu_0}{2}\frac{\partial \tau}{\partial U_0}
            \right)
        },
        \label{eq:growth rate formula}%
\end{equation}
%
%where henceforth, for notational expedience, we write $\frac{\partial \tau}{\partial
%U_0}\equiv \big(\frac{\partial \tau}{\partial U}\big)_0$ and so on, for
%derivatives evaluated at the base state.
The system is unstable when the real part of this formula is positive.
For unconfined flows, both branches of $\real(\sigma)$
are even functions of $k$ and monotonic in $|k|$~\citep{Langham2022}.  Identical
reasoning applies in this more general case.  Consequently, for each mode,
$\real(\sigma)$ is bounded by its value at $k=0$ and in the limit $k\to \infty$.

At zero wavenumber, $\sigma = 0$ or $-\partial\tau/\partial U_0$.  Therefore,
for drag laws
with $\partial\tau/\partial U_0 < 0$, such as within the hysteretic parts of
some depth-averaged granular friction models~\citep{Edwards2019}, the flow is
unconditionally unstable.
More commonly, $\partial\tau/\partial U_0 > 0$ and the presence of instability 
is dictated by the sign of $\real(\sigma)$ in the high wavenumber regime.
This shall be assumed to be the case hereafter.
By decomposing into real and imaginary parts and taking the limit of the
resulting expressions,
it may be determined that
%%
%\begin{equation}
%    \real(\sigma) \to \frac{1}{2}
%    \left[
%        -\frac{\partial \tau}{\partial U_0}
%        \pm
%        \Fr
%        \left(
%        \frac{1}{\Delta}\frac{\partial \tau}{\partial s_0} - 1
%        \right)
%        \right]
%\end{equation} 
%%
\begin{equation}
    \real(\sigma) \to \frac{1}{2}
    \left[
        -\frac{\partial \tau}{\partial U_0}
        \pm
        \frac{\Fr}{\sqrt{1 + \Fr^2 (\shapefu_0^2/4 + \shapefs_0)}}
        \left(
        \frac{1}{\Delta}\frac{\partial\tau}{\partial s_0}
        -1
        + \frac{\shapefu_0}{2}\frac{\partial\tau}{\partial U_0}
        \right)
        \right]
    \label{eq:asym growth}%
\end{equation}
as $|k|\to\infty$. When drag is mediated via stresses arising from velocity
shear, it
might be anticipated that $\partial \tau / \partial s_0 < 0$, since
increasing the surface elevation ought to decrease these
stresses at the bed. 
%Indeed, as demonstrated below, even drag models that are
%not shear dependent can have this feature.  
Moreover, for the normal-stress dependent granular drag law considered below
(in \S\ref{sec:granular examples}),
$\Delta^{-1}\partial\tau/\partial s_0 - 1 < 0$.  In such cases, an implication
of this formula is that when both $\alpha_0 \geq 0$ and $\beta_0\geq 0$ (which
is true for any constant $\chi$, as well as the more realistic shape factor
formulae derived later on), only the negative branch of~\eqref{eq:asym growth}
can be unstable, and the effect of increasing the shape factor above unity
decreases the growth rate of this branch.  In general,
confined channel flows must possess larger shape factors than the
unconfined case if there is no slip at the channel base, since their velocity
profiles necessarily feature cross-stream variation.
%If a constant shape factor function is then assumed,
%for increases of $\shape_0$ above~$1$, the presence of
%the shape factor reduces growth rates, provided that
%$(\shape_0-1)\partial\tau/\partial U_0$ does not become so large that the positive
%branch becomes the dominant mode.
%As we shall we in \S\ref{sec:examples}, in practice even simple models of $\chi$
%acquire some dependence on $s$.
%Though
%it is tempting to further conclude that the geometric parameter $\Delta$
%necessarily reduces the growth rate of the dominant mode when $\Delta > 1$,
%caution is warranted at this stage, since geometric information is likely to be
%embedded within $\tau$ also. This will become apparent when specific examples
%are presented in~\S\ref{sec:examples}.
It is tempting to conclude further from~\eqref{eq:asym growth} that the
geometric information encoded in $\Delta$ implies that all convex open channels
are stabilising (relative to unconfined flows), on the grounds that the negative
branch of $\real(\sigma)$ is a decreasing function of $\Delta$, if all other
terms are held fixed.  While this intuition is borne out in our later analysis,
a fully general statement of this kind is not possible, since both $\tau$ and
$\chi$ can also depend on the channel shape.

One of the asymptotic growth rates in~\eqref{eq:asym growth} is
strictly positive whenever
%
%\begin{equation}
%    \Fr > 
%    \frac{\frac{\partial \tau}{\partial U_0}}{\left|1 -
%    \frac{1}{\Delta}\frac{\partial \tau}{\partial s_0}\right|}.
%\end{equation}
%
\begin{equation}
    \Fr^2\left[
    \left(1 - \frac{1}{\Delta}\frac{\partial\tau}{\partial s_0}
    -\frac{\shapefu_0}{2}\frac{\partial\tau}{\partial U_0}
    \right)^2
    -(\shapefu_0^2/4 + \shapefs_0)\left(
    \frac{\partial\tau}{\partial U_0}
    \right)^2
        \right]
    > \left( \frac{\partial \tau}{\partial U_0}\right)^2\!\!\!.
    \label{eq:instab ineq}%
\end{equation}
At this point, we turn to the steady uniform balance~\eqref{eq:steady
balance}, which
in dimensionless variables is $\tau(A, U) = A$.
Using the implicit function theorem, this expression may be parameterised in
terms of $s$. That is, we write $\tau(s, U(s)) = A(s)$, where $U(s)$ gives the
steady velocity for a given surface elevation.
On differentiating this with respect to $s$ and evaluating at the base state,
one obtains
\begin{equation}
    \frac{\partial \tau}{\partial s_0} + \frac{\partial \tau}{\partial U_0} U_0'
    = \Delta,
    \label{eq:steady balance diff}%
\end{equation}
with the prime henceforth denoting (total) differentiation with respect to $s$.
We use~\eqref{eq:steady balance diff} to eliminate $\partial \tau/\partial U_0$
in~\eqref{eq:instab ineq}. Then, considering the conditions for
which~\eqref{eq:instab ineq} may be satisfied, we conclude that instability
occurs if
%If $\shape_0 = 1$, the expression within the square brackets is guaranteed to be
%positive. However, if  $\shape_0 > 1$, it can be negative, making~\eqref{eq:instab
%ineq} impossible to satisfy.  From this observation, a threshold can be obtained
%for $\shape_0$ beyond which the flow is stable for all $\Fr$.
%Combining this with~\eqref{eq:instab ineq}, 
%Hence, for instability to occur, it must be the case that 
%$\Fr > \Fr_c$, with
%$\Fr_c\equiv \Fr_c(\alpha_0,\beta_0)$ defined by
%
\begin{subequations}
\begin{equation}
    \Fr>\Fr_{c} = \frac{1}{\left| \left( 
    \frac{U_0'}{\Delta} - \frac{\shapefu_0}{2}
    \right)^2
    -\frac{\shapefu_0^2}{4} - \shapefs_0
    \right|^{1/2}}
    ~~
    \mathrm{and}
    ~~
    \shape_0 \ngtr
    1 + 
    \frac{1}{2 + \Delta/U_0'}\left[
        \frac{U_0'}{\Delta} - \frac{\partial \shape}{\partial U_0} -
        \frac{1}{U_0'}\frac{\partial \shape}{\partial s_0}
        \right]\!,
    %\frac{(C_0'/\Delta - 1)^2}{2C_0'/\Delta - 1}.
    \tag{\theequation\emph{a,b}}%
    \label{eq:sectional trowbridge}%
\end{equation}
    \label{eq:sectional trowbridge single}%
\end{subequations}
with $\Fr_c$ denoting the critical Froude number.
The right-hand inequality is obtained from the requirement that the left-hand
side of~\eqref{eq:instab ineq} be positive
and referring back to~\eqref{eq:alpha beta}.
In the case where shape factors are neglected, 
$\shapefu_0 = \shapefs_0 = 0$ and~\eqref{eq:sectional trowbridge} reduces to
\begin{equation}
    \Fr_c = \frac{\Delta}{|U_0'|}.
    \label{eq:sectional trowbridge simple}%
\end{equation}
This pleasingly simple expression is a generalisation of similar formulae for
unconfined channels \citep[see][]{Craya1952,Langham2022}.

%\begin{subequations}
%\begin{equation}
%    \Fr > \Fr_c,\quad\mathrm{with}\quad
%\Fr_c(\shapefu_0, \shapefs_0) =
%    \frac{\frac{\partial\tau}{\partial U_0}}{\left|
%    \left(1 - \frac{1}{\Delta}\frac{\partial\tau}{\partial s_0}
%    -\frac{\shapefu_0}{2}\frac{\partial\tau}{\partial U_0}
%    \right)^2
%    -\left(\frac{\shapefu_0^2}{4}+\shapefs_0\right)\left(
%    \frac{\partial\tau}{\partial U_0}
%    \right)^2
%    \right|^{1/2}}.
%        \label{eq:gen trowbridge 1}%
%\end{equation}
%
%Additionally, the left-hand side of inequality~\eqref{eq:instab ineq} must be
%strictly positive. This requirement can be written as
%%
%\begin{equation}
%    \shape_0 \ngtr 1 + \frac{1}{\Fr_c(0,0)+2}\left(
%    \frac{1}{\Fr_c(0,0)} - \frac{\partial \shape}{\partial U_0} - \Fr_c(0,0)\frac{\partial
%    \shape}{\partial s_0}
%        \right).
%        \label{eq:gen trowbridge 2}%
%\end{equation}
%    \label{eq:gen trowbridge}%
%\end{subequations}
%
The~\eqref{eq:sectional trowbridge} conditions are
necessary and sufficient for the existence of a linear instability.
Physically, this can be thought of as arising when resistive forces at the
flow base fail to dissipate the energy provided to small wavelength
disturbances by gravity, causing them to grow.
This intuition is made concrete in Appendix~\ref{appendix:energy stability} for
the case of $\chi \equiv 1$, with a brief analysis of the energy equation.
In principle, via specification of $\tau$, $\Delta$ and
$\shape$,~\eqref{eq:sectional trowbridge}
captures the linear stability threshold for many different flow configurations.
However, this must be tempered by the knowledge that roll wave instabilities are
convective in nature~\citep[see e.g.][]{Forterre2003}, which raises the
possibility that unstable modes can propagate out of a finite channel before
they grow to reach observable amplitudes.

The modes grow in space with rate $\real(\sigma)/c$,
and it may be verified that this expression
is likewise maximised in the high
wavenumber limit. In this regime, the phase speeds are equal to the
characteristics of the system, which are the eigenvalues of the Jacobian of
equations~\eqref{eq:governing nondim 1} and~\eqref{eq:governing nondim 2}. We
compute
\begin{equation}
    c \to 1 + \frac{\shapefu_0}{2} \mp \frac{1}{\Fr}\sqrt{1+\Fr^2(\shapefu_0^2/4
    + \shapefs_0)},
    \label{eq:c}%
\end{equation}
as $|k|\to\infty$. The two branches are signed to match the corresponding
formula in~\eqref{eq:asym growth}. Therefore, following our discussion from
above, it is typically the positive branch that gives the speed of unstable
modes, if present. 
While the non-dimensionalisation in~\eqref{eq:nondimensionalisation} is
convenient for calculations, it is better to compare the spatial growth rate
against the channel length $\tilde L$. Their ratio
asymptotes to
\begin{equation}
    \frac{\real(\tilde \sigma) \tilde L}{\tilde c} \to \frac{\real(\sigma) \tilde L
    \tan\theta}{\Fr\left[\Fr(1+\alpha_0/2) \mp \sqrt{1 +
    \Fr^2(\shapefu_0^2/4+\shapefs_0)}\right]\tilde D_0},
    \label{eq:asym spatial growth}%
\end{equation}
as $|k| \to \infty$ (where $\tilde c = c \tilde{U}_0$ and $\tilde \sigma =
\sigma\tilde{U_0}/\tilde{\ell}$).  As $\Fr$ becomes very large in this
expression, the denominator scales like $O(\Fr^2)$.  If neither $\tau$, nor
$\chi$ depend on $\Fr$, then $\real(\sigma)$ grows more slowly according
to~\eqref{eq:asym growth}: $O(\Fr)$ if $\shape_0 = 1$ or $O(1)$ if $\shape_0 >
1$.  Furthermore, since increasing the hydraulic depth in steady conditions
necessitates a corresponding increase in velocity to maintain the same shear
profile, $\tilde D_0$ is expected to be an increasing function of $\Fr$.
Consequently, for a fixed channel,~\eqref{eq:asym spatial growth} implies that
the spatial growth rate ultimately decays as $\Fr \to \infty$ and must thereby
be maximised at some intermediate value for unstable cases.  This property has
previously been noticed in experiments and models of unconfined granular
flows~\citep{Forterre2003}.  However, in this particular case the basal drag does
depend on $\Fr$ and $\Real(\sigma) = O(1)$ when $\chi_0 = 1$, or $O(\Fr^{-1})$
when $\chi_0 > 1$.

\section{Illustrative special cases}
\label{sec:examples}%
Some further assumptions are useful to develop an understanding of
general analyses conducted above. 
In the following subsections, we make a progressive sequence of assumptions
about the system that demonstrate how to apply the theory and draw out some
practical implications.

\subsection{Laterally shallow flows}
\label{sec:shallow flows}%
A particularly powerful simplification is to
assume that the flow depth $\tilde h$ is typically far smaller than the width of
the wetted region, which is often the case for observations of roll
waves~\citep[e.g.][]{Depoilly2024,Chen2024,Aaron2025}.
Though this implies that the flow is only rather weakly confined, 
we demonstrate below that this is enough to significantly alter its
stability.
In this case, the section-averaged governing equations are
equivalent to separately width-averaging a depth-averaged flow with
(dimensional) downstream velocity $\bar{u}(\tilde x, \tilde y)$. 
A derivation of this fact is presented in Appendix~\ref{appendix:width averaging}.
%Given this, the functional dependence of the
%aggregate drag on the cross-section may be obtained with the following argument.
This allows us to leverage expressions for depth-averaged quantities in the
formulae derived above.

Integrating the depth-averaged velocity over the section gives the flux $\tilde
Q$ as
\begin{equation}
    %\tilde Q = 
    \tilde{Q} = \tilde{A}\tilde{U} = \int_{-\infty}^{\infty}
    \bar{u}\tilde{h}\,\mathrm{d}\tilde{y}.
    \label{eq:flux}%
\end{equation}
Then, after defining $Q = \tilde Q / \tilde Q_0$, we note that $U_0' = Q_0' -
\Delta$. This expression, which is needed to evaluate the critical Froude
number~\eqref{eq:sectional trowbridge}, may be computed exactly 
for a particular flow
by using~\eqref{eq:flux}, with 
$\bar{u}$ replaced by the function $\bar{u}_0$ that locally parametrises the
depth-averaged velocity under steady uniform conditions, in terms of the flow
depth.
%The depth-averaged downstream velocity $\bar{u}_0$ in steady uniform flow
%conditions is given by
%%
%\begin{equation}
%    \bar{u}_0 = %\tilde K \tilde h^p
%    \mathcal{U}(\tilde h_0),
%    \label{eq:1d power law}%
%\end{equation}
%%
%where $\mathcal{U}$ is some function dictated by the steady balance, 
%that can be prescribed, or determined empirically.
For example, $\bar{u}_0$ is frequently expressible as a power law,
$\bar{u}_0(\tilde h) = \tilde K\tilde h^p$, with $\tilde K$ independent of
$\tilde h$ and $p = 2$ for a laminar Newtonian fluid~\citep{Yih1963}, $p = 3/2$ for dense dry
granular flows of smooth monodisperse spheres~\citep{Pouliquen1999} and $p = 1/2$ for the Ch\'ezy
model of turbulent drag~\citep{Balmforth2004b}.  
In such cases, many aspects of the stability problem depend upon integrals of
the flow depth, raised to some power, across the section.
Because of this, it is convenient to define the functionals
\begin{equation}
    \mathcal{J}(n) = \frac{1}{\tilde{s}_0^{n+1}}
    \int_{-\infty}^{\infty}\tilde{h}^n\,\mathrm{d}\tilde{y}.
    \label{eq:J(n)}%
\end{equation}
On differentiating with respect to $s$, it may be verified that
$\mathcal{J}'(n) = (n+1)\mathcal{J}(n)$.  Using this, formulae for the main
ingredients of the stability problem may be deduced in terms of these
functionals:
\begin{subequations}
\begin{equation}
    \Delta = \frac{\mathcal{J}_0(0)}{\mathcal{J}_0(1)},
    \quad
    Q_0' = \frac{(p+1)\mathcal{J}_0(p)}{\mathcal{J}_0(p+1)},
    \tag{\theequation\emph{a,b}}%
    \label{eq:Delta Q0p power law}%
\end{equation}
    \label{eq:Delta Q0p power law single}%
\end{subequations}
where the latter expression has been obtained via~\eqref{eq:flux} with
$\bar{u}_0(\tilde{h}) = \tilde{K}\tilde{h}^p$.

To evaluate the most general form of the $\Fr_c$ curve, a formula for the shape
factor $\chi$~\eqref{eq:shape factor} is needed.  In general, this depends on
the local unsteady three-dimensional flow field, which cannot be calculated in
this setting.  However, for the stability problem, $\chi$ and its derivatives
need only be evaluated at equilibrium.  Therefore, we compute~\eqref{eq:shape
factor} by assuming that at each point along the channel cross-section, the flow
adopts a steady velocity profile $\tilde{u}_0(\tilde y, \tilde z)$, whose
corresponding depth-averaged value is $\bar{u}_0(\tilde{h})$.
This leads to
\begin{equation}
    \shape = \frac{\tilde A}{\tilde Q^2}\int_{-\infty}^\infty
    \chi_\infty
    \bar{u}_0(\tilde h)^2 \tilde h\,\diff \tilde y,
    \quad
    \mathrm{where}
    \quad
    \shape_\infty \equiv \frac{1}{\tilde h \bar{u}^2}\int_{0}^{\tilde h}
    \tilde u_0^2(\tilde y, \tilde z - \tilde b)\,\diff \tilde z.
    \label{eq:shallow shape factors}%
\end{equation}
Given the lateral shallowness assumption,
the functional form of $\tilde u_0(\tilde y,
\tilde z - \tilde b)$ can be reasonably anticipated to be independent of $\tilde
y$, implying that $\chi_\infty$ is given by the shape factor for the
corresponding unconfined flow. That is, $\chi_\infty$ accounts for velocity
shear across the depth. Since this is a dimensionless quantity, it cannot itself
vary with the absolute flow depth unless $\tilde u_0$ depends on another length scale.
Consequently, $\chi_\infty$ is often a constant that can be factored out of the
integrand in the formula for $\chi$.

If this is the case, and if $\bar{u}_0 = \tilde K \tilde h^p$, then $\chi$ may
be expressed as
\begin{equation}
    \chi(s) = \chi_\infty
    \frac{\mathcal{J}(1)\mathcal{J}(2p+1)}{\mathcal{J}(p+1)^2}.
    \label{eq:chi power law}%
\end{equation}
The formulae in~\eqref{eq:Delta Q0p power law} and~\eqref{eq:chi power law} may
be used in~\eqref{eq:sectional trowbridge} to compute the critical Froude number
for flow in a particular channel by determining $\mathcal{J}(n)$.
Vertical shear may be neglected, independently from cross-stream dependence, by
setting $\chi_\infty = 1$.

%Alternatively, in the case where shape factors are neglected, 
%$\shapefu_0 = \shapefs_0 = 0$ and~\eqref{eq:sectional trowbridge} becomes
%%
%\begin{equation}
%    \Fr_c = \frac{\Delta}{|U_0'|}.
%    \label{eq:sectional trowbridge simple}%
%\end{equation}
%%
%This pleasingly simple expression is a generalisation of similar formulae for
%unconfined channels \citep[see][]{Craya1952,Langham2022}.

\subsection{Trapezoidal channels}
\label{sec:trapezoids}%
Therefore, to progress further, it is necessary to select a geometry.  In this
subsection, we specialise to the case of a symmetric convex trapezoidal channel
with base width $\tilde w$ and banks of gradient $S_b$.
The case of channel cross sections defined by
a power law dependence is also tractable, but more mathematically complicated,
and leads to very similar conclusions. We show how to work through this in
Appendix~\ref{appendix:power law}.

Hereafter, the width of the trapezoidal channel base 
is rendered dimensionless by the steady free
surface elevation, by defining $w = \tilde w / \tilde s_0$.  This geometry is
simple enough to permit typical calculations to be evaluated analytically, while
the parameters $w$ and $S_b$ can be continuously adjusted to roughly approximate
a wide range of natural and man-made channels.  In the limit $w\to \infty$, the
channel becomes `hydraulically wide' and the expressions we derive below tend to
the corresponding formulae for uni-dimensional depth-averaged models.  The $w =
0$ case gives the opposite extreme of a triangular channel.  Likewise, the $S_b
\to 0$ limit with $w > 0$ held fixed, is also `triangular', at least from the
perspective of this theoretical framework.  To see why, consider a sequence of
channels with $S_b$ decreasing. As the banks become shallower, the relative
fraction of the wetted basal perimeter that lies along these angled regions
increases, until the presence of the flat basal section is rendered negligible
by comparison.

\begin{figure}
    \begin{center}
    \includegraphics[width=\columnwidth]{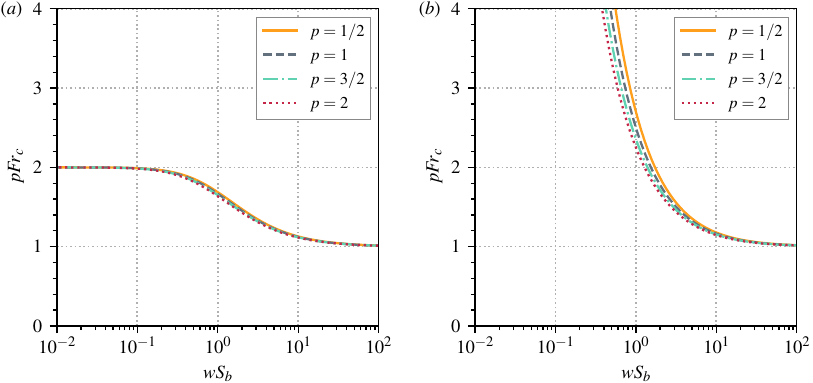}%
    \end{center}
    \caption{%
        Ratio of critical Froude number $\Fr_c(wS_b)$ to its value in the
        unconfined limit ($\Fr_c = 1/p$) for a trapezoidal channel and fluids
        with power-law dependence $p = 1/2$ (Ch\'ezy, solid yellow), $1$ (dashed
        grey), $3/2$ (granular, dotted red) and $2$ (viscous, dash--dotted
        turquoise).  Panels (a) and (b) show different choices for the shape
        factor: (a)~$\shape = 1$ and (b)~$\shape$ set by Eq.~(\ref{eq:trap geom
        params}c) with $\shape_\infty = 1$.
    }
    \label{fig:Frcrit}%
\end{figure}

For this geometry, the integrals in~\eqref{eq:J(n)} evaluate as
\begin{equation}
    \mathcal{J}(n) = ws^n + \frac{2s^{n+1}}{(n+1)S_b}.
    \label{eq:J(n) trap}%
\end{equation}
Substituting this into~\eqref{eq:Delta Q0p power law}, \eqref{eq:chi power law} and
simplifying leads to general formulae that depend only upon $p$ and $wS_b$:
%In the case of the trapezoidal section, these are not mutually independent and
%depend only on $p$ and $wS_b$.
%Elementary calculations using the geometry of the trapezium lead to the
%following general formulae:
%
\begin{subequations}
\begin{equation}
    \Delta = \frac{wS_b + 2}{wS_b + 1},\quad Q_0' = \frac{(p+1)wS_b +
    2}{wS_b + \frac{2}{p+2}},\quad
    \shape(s) = \shape_\infty\frac{(wS_b + s)\left(wS_b +
    \frac{s}{p+1}\right)}{\left(wS_b + \frac{2s}{p+2}\right)^{\!2}}.
    \tag{\theequation\emph{a--c}}%
    \label{eq:trap geom params all}%
\end{equation}
    \label{eq:trap geom params}%
\end{subequations}
Since shape factors are often neglected, either for model
simplicity~\citep[e.g.][]{Iverson1997,Fraccarollo2000,Aaron2025}
or because a plug-like profile is anticipated~\citep[e.g.][]{Balmforth2004b},
we first consider the
simpler case of $\shape = 1$.
%
%By computing $C_0'/\Delta$ and considering the relative sizes of the terms on
%the numerator and denominator, it may be shown that $C_0' > \Delta$, provided
%that $p > 0$ and $wS_b \geq 0$.
On substituting $\Delta$ and $U_0' = Q_0'-\Delta$ into~\eqref{eq:sectional
trowbridge simple}, 
the critical Froude number for this case is found to be
\begin{equation}
    \Fr_{c} =
    \frac{(p+2)(wS_b)^2 + 2(p+3)wS_b + 4}
    {p(p+2)(w S_b)^2 + p(p+3)wS_b + 2p}.
    \label{eq:Frcrit beta 1}%
\end{equation}
This formula recovers the unconfined critical value $\Fr_{c} = 1/p$ in the
corresponding limit $wS_b\to\infty$~\citep{Langham2022}.  
In the other extreme, $\Fr_{c} =
2/p$ for a triangular channel with $wS_b = 0$. 
By 
consulting~(\ref{eq:trap
geom params}\emph{a,b}), 
one sees that $U_0' = p$ in either
of these cases, so the doubling of $\Fr_c$ as the triangular channel is
approached can be attributed to the doubling of
the geometric parameter $\Delta$.
%Since the relative change in the critical Froude number is the same at either
%extreme, the $\Fr_c$ curves approximately collapse under rescaling by $1/p$.
In figure~\ref{fig:Frcrit}(\emph{a}), we plot the ratio of $\Fr_c$ to the
limiting value in the unconfined case for $p = 1/2,1,3/2$ and $2$.
For these modest (but realistic) values of $p$, the curves approximately
collapse onto a single line.
%Furthermore, we see that for any $1 < p\Fr_c < 2$, confinement ultimately
%stabilises the flow.

Turning to the case where the momentum shape factor of~(\ref{eq:trap geom
params}\emph{c}) is included, 
%a choice must be made for the value of $\shape_\infty$.
%This depends on problem-specific features of the vertical velocity profile that
%are independent of the assumptions made thus far.  Therefore, we first cover the
%simplest case, where $\shape_\infty = 1$, which corresponds to the common
%decision to neglect vertical velocity shear.  
if the simplifying assumption $\shape_\infty = 1$ is employed, then
by using the expressions of~\eqref{eq:trap geom params all} in~\eqref{eq:sectional trowbridge},
the stability boundary is
\begin{equation}
    \Fr_c = 
    \frac{(p+2)(wS_b)^2 + 2(p+3)wS_b + 4}
    {p(p+2)wS_b(wS_b + 1)}.
    \label{eq:Frc chiinf = 1}%
\end{equation}
As $wS_b \to \infty$, this recovers the unconfined value $\Fr_c \to 1/p$ (as it
must).  Furthermore, for $wS_b \ll 1$,
\begin{equation}
    p\Fr_c = \frac{4}{(p+2)wS_b} + O(1),
    \label{eq:pFrc divergence}%
\end{equation}
indicating that the triangular 
geometry is unconditionally stable in this case.  Example $p\Fr_c$ curves are
plotted in figure~\ref{fig:Frcrit}(\emph{b}) to demonstrate these features.  By
comparison with the $\shape = 1$ stability boundaries of
figure~\ref{fig:Frcrit}(\emph{a}), we see that the effect of the cross-stream velocity
profile, as captured by the shape factor with $\shape_\infty = 1$, becomes
prominent when $wS_b \lesssim 10$ and is strongly stabilising.  

The same calculations that led to~\eqref{eq:Frc chiinf = 1} may be repeated
for a general vertical shape factor $\shape_\infty \geq 1$, 
ultimately leading to 
\begin{equation}
    \Fr_c =
    \frac{(p+2)(wS_b)^2 + 2(p+3)wS_b + 4}
    {(p+2)(w S_b + 1)\sqrt{(wS_b)^2p^2 - 
    (\shape_\infty-1)[(w S_b)^2(2p+1) + 4 wS_b(p+1) + 4]}}.
    %{(p+2)w S_b (w S_b + 1)\sqrt{(wS_b)^2[p^2 - (2p+1)(\shape_\infty-1)]
    %-4(\shape_\infty-1)(w S_b(p+1)+1)}}.
    %{p(p+2)w S_b (w S_b + 1)}.
%    \frac{2 + wS_b}{p(p+2)wS_b}
%    \sqrt{\frac{
%        (wS_b(p + 2) + 2)^3(p + 1)}{
%        (p+1)[(p+2)(wS_b)^3 + 2(p+3)(wS_b)^2] + (p^2 + 7p + 5)wS_b + 2p}}.
%        %(p^2 + 3p + 2)(wS_b)^3 + (2p^2 + 8p + 6)(wS_b)^2 + (p^2 + 7p + 5)wS_b + 2p}}.
    \label{eq:Frcrit beta cross}%
\end{equation}
which %, for modest values of $\shape_\infty$, 
is applicable over the parameter ranges
\begin{subequations}
\begin{equation}
    \frac{w S_b}{2} > \frac{(p+1)(\shape_\infty - 1) +
    p\sqrt{\shape_\infty(\shape_\infty-1)}}{p^2 - (\shape_\infty-1)(2p+1)}
    \quad\mathrm{and}\quad
    1 \leq \shape_\infty < 1 + \frac{p^2}{2p+1}.
    \tag{\theequation\emph{a,b}}%
    \label{eq:Frcrit conditions both}%
\end{equation}
    \label{eq:Frcrit conditions}%
\end{subequations}
While lengthy, these explicit expressions enable us to make some general
observations about this case.  Provided the~(\ref{eq:Frcrit conditions
both}) conditions are satisfied,~\eqref{eq:Frcrit beta cross} is a
continuous, differentiable curve and strictly decreasing as a function of
$wS_b$.  At the limit of inequality~(\ref{eq:Frcrit conditions}\emph{a}) it features a
singularity and for lower, yet still positive $w S_b$ values, the flow is
unconditionally stable.  An example $\Fr_c$ curve, demonstrating this property is
plotted in the next subsection (figure~\ref{fig:spatial growth
granular}(\emph{c})).
Fixing $p$ and increasing $\shape_\infty$ broadens this stable region until the
upper limit of~(\ref{eq:Frcrit conditions}\emph{b}) is reached.  Beyond this point,
different conditions for $\Fr_c$ apply, which bound the stability boundary to a
finite open interval. However, since $\shape_\infty$ is usually near unity, this
likely lies outside the physical applicability of the theory for most fluids.
For example, the vertical shape factor of a power law fluid is $\shape_\infty =
1 + 1/(2p+1)$ with $p>1$~\citep{Ng1994}, which trivially
satisfies~(\ref{eq:Frcrit conditions}\emph{b}).

\subsection{Tilted trapezoids}
\label{sec:tilted trapezoids}
\begin{figure}
    \begin{center}
        \includegraphics[width=9.5cm]{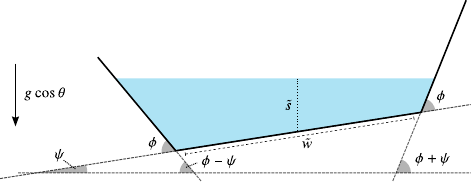}%
    \end{center}
    \caption{Diagram of the tilted trapezoidal channel cross section
    showing a geometric construction that may be used to determine
    $\mathcal{J}(n)$.
    The downslope direction is orthogonal to the page in this schematic.}
    \label{fig:tilt schematic}%
\end{figure}

Though laboratory chute flows are generally designed to be level in the
cross-stream direction, small amounts of bias can be unwittingly introduced.
This is often easily detected visually, by observing asymmetries in either the
flow front, the velocity profile, or any waves that develop.  A  large-scale
example is provided by the US Geological Survey's $95\,\mathrm{m}$-long debris
flow flume experiment, which suffers from roughly $2^\circ$ lateral tilt
\citep{Iverson2010}.
Likewise, natural channels are rarely perfectly symmetric with a flat
base and may posses deeper regions towards one side or the other.
To explore the effect this has on the roll wave instability, we consider the
channel geometry depicted in figure~\ref{fig:tilt schematic} -- a trapezium with
bank angle $\phi$, tilted at an angle $\psi$ to the horizontal, with $\tilde{s}$
measured from the centre of the trapezoidal base.

Like~\eqref{eq:J(n) trap}, $\mathcal{J}(n)$ can be determined using elementary
geometry, but the resulting expression is more complicated due to the asymmetry.
By partitioning the integral into a trapezium, plus two triangles at the banks,
one obtains
\begin{equation}
    \mathcal{J}(n) = 
    \frac{s^{n+1}}{n+1}\left\{
        \frac{1}{S_1}\left(1 - \frac{\delta}{2s}\right)^{n+1}
    \!\!+\frac{1}{S_2}\left[
         \left(
        1 + \frac{\delta}{2s}
        \right)^{n+1}
         - \left(
        1 - \frac{\delta}{2s}
        \right)^{n+1}
        \right]
    +
        \frac{1}{S_3}\left(1 + \frac{\delta}{2s}
    \right)^{n+1}
    \right\}\!,
\label{eq:J(n) tilt}%
\end{equation}
with $\delta = w\sin\psi$, $S_1 = \tan(\phi + \psi)$, $S_2 = \tan(\psi)$, and
$S_3 = \tan(\phi - \psi)$.  This expression is valid for values of the tilt
angle, up to the point where either the bank gradients change sign, or the
wetted region becomes triangular. Therefore, we limit $\psi$ to
\begin{equation}
    |\psi| \leq \min\{\phi, \pi/2-\phi, \arcsin(2s/w)\}.
    \label{eq:psi condition}%
\end{equation}
Note that~\eqref{eq:J(n) tilt} and~\eqref{eq:psi condition} are straightforward
to extend to trapezia with two different bank angles, if needed.

\begin{figure}
    \begin{center}
    \includegraphics[width=\textwidth]{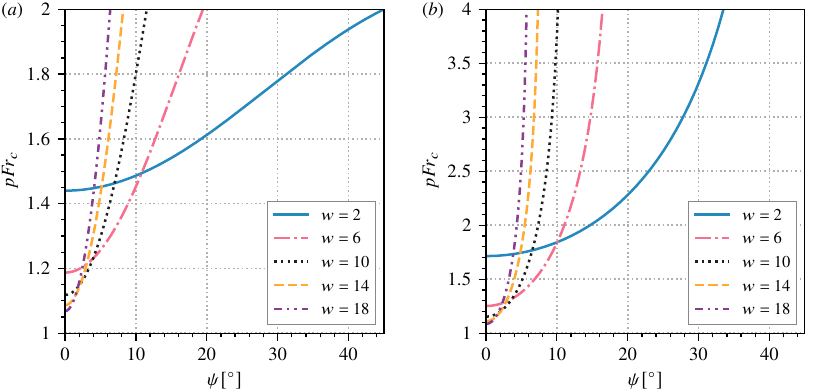}%
    \end{center}
    \caption{Critical Froude numbers for trapezoidal channels tilted at an angle
    $\psi$ to the horizontal, with different dimensionless base widths $w =
    \tilde w / \tilde{s}_0$.  
    The bank angle is $\phi = 45^\circ$ and $p=3/2$.
    As in figure~\ref{fig:Frcrit}, the vertical axis is rescaled by
    $\Fr_c$ in the unconfined limit ($\Fr_c = 1/p$).
    Panel~(\emph{a}) shows the case where the shape factor is set to unity,
    while~(\emph{b}) uses the shape factor given in~\eqref{eq:chi power law} with 
    $\chi_\infty = 1$.
    Note that, in~(\emph{a}) 
    the full extent of each curve is included, within the range set
    by~\eqref{eq:psi condition}, whereas for~(\emph{b}) the curves asymptote to
    infinity as $\psi$ reaches the limits imposed by~\eqref{eq:psi condition},
    so only part of their vertical extent is shown.
    }
    \label{fig:tilt}%
\end{figure}

Curves for the critical Froude number may be obtained as before, 
by using $\mathcal{J}(n)$ in~\eqref{eq:Delta Q0p power law},~\eqref{eq:chi power
law} and substituting the resulting expressions into~\eqref{eq:sectional
trowbridge}. We do so numerically to produce the curves in
figure~\ref{fig:tilt}.
In this variant of the problem, the results do not depend neatly on only one
geometric parameter. However, they are qualitatively insensitive to the bank
angle and steady velocity exponent, so we plot $\Fr_c$ for $\phi = 45^\circ$,
$p=3/2$ as a representative example, and show different choices of $w =
\tilde{w}/\tilde{s}_0$. 
We see that as $\psi$ increases, so does $\Fr_c$.  This is to be expected, since
the wetted region approaches a triangle as the channel becomes more tilted. For
example, in figure~\ref{fig:tilt}(\emph{a}), which shows the curves with shape
factors neglected, $\Fr_c$ reaches $2/p$ at the uppermost $\psi$ values, just as
in the symmetric case without tilt, plotted in
figure~\ref{fig:Frcrit}(\emph{a}).  When shape factors are included, as in
figure~\ref{fig:tilt}(\emph{b}), the stabilising effect of channel tilt is more
severe, and $\Fr_c\to\infty$ as the wetted area approaches a triangle, which may
be compared to the figure~\ref{fig:Frcrit}(\emph{b}) curves.  In both cases,
wider channels are more sensitive to variations in $\psi$, since the angle
between the untilted case and the triangular case is smaller.
In particular, it is notable that only a handful of degrees are needed to fully
stabilise the channel in figure~\ref{fig:Frcrit}(\emph{b}) for the largest $w$
values plotted.

\subsection{Granular friction}
\label{sec:granular examples}%
The conclusions made so far have been qualitatively independent of the choice
of shallow rheological model. 
We now focus on the case of dense, dry flows of monodisperse particles, which can be 
a good model for natural gravity currents that feature high concentrations of
granular material~\citep{Aaron2025}.
%To this end, we set $\mathcal{F}$ so that the corresponding drag
%model captures a section-averaged granular
%friction law, derived in Appendix~\ref{appendix:granular drag}:
In steady uniform conditions, the depth-averaged velocity $\bar{u}$ of a monodisperse
granular flow in a hydraulically wide channel
is well captured by the relation
\begin{equation}
    \bar{u} = \sqrt{g\cos\theta}\left(
    \frac{\xi \tilde h^{3/2}}{\tilde{h}_{\mathit{stop}}(\theta)} - 
    \gamma \tilde h^{1/2}
    \right),
    \label{eq:ubar grains}%
\end{equation}
where $\xi$ and $\gamma$ are positive dimensionless constants and
$\tilde{h}_{\mathit{stop}}(\theta)$ is the measured deposit thickness left by
such a flow after it subsides~\citep{Pouliquen1999,Forterre2003}.  This
expression forms the primary basis for models that capture the basal friction of
shallow dense granular flows, which are typically constructed to capture all
possible steady states across a finite range of slope
angles~\citep{Pouliquen2002,Edwards2019}.  When $\gamma = 0$, \eqref{eq:ubar
grains} is a consequence of the well-known Bagnold scaling law for flows of spherical
particles~\citep{GDR2004}, while $\gamma > 0$ is a phenomenological correction
that is practically useful for modelling angular particles, which arrest more
readily than spherical ones~\citep{Forterre2003}.  Multiple experimental
studies~\citep{Pouliquen1999,Pouliquen2002,Forterre2003,Edwards2017} have
separately concluded that
\begin{equation}
    \tilde{h}_{\mathit{stop}}(\theta) = 
    \tilde{\mathcal{L}}\left(
    \frac{\tan\theta_d - \tan\theta}{\tan\theta - \tan\theta_s}
    \right),
    \label{eq:hstop}%
\end{equation}
where $\tilde{\mathcal{L}}$ is a characteristic length scale for the grains and
$\theta_s$, $\theta_d$ are the minimum and maximum slope angles respectively,
for which steady flows are observed.

As alluded to previously, the case $\gamma = 0$ in~\eqref{eq:ubar grains}
corresponds to setting $p=3/2$ in the analysis of \S\ref{sec:shallow
flows}--\ref{sec:tilted trapezoids}.  The more general velocity relation
of~\eqref{eq:ubar grains} means that $Q_0'$ must be rederived.
Integrating~\eqref{eq:ubar grains} over the channel cross-section gives
\begin{equation}
    \tilde U = \frac{\sqrt{g\cos\theta}}{\tilde A}\left(
    \frac{\xi}{\tilde{h}_{\mathit{stop}}} 
    \int_{-\infty}^{\infty}\tilde h^{5/2}\,\mathrm{d}\tilde{y} - 
    \gamma 
    \int_{-\infty}^{\infty}\tilde h^{3/2}\,\mathrm{d}\tilde{y}
    \right)\!.
    \label{eq:U grains}%
\end{equation}
From this, we construct a function for the basal coefficient of friction $\mu$ in the
section-averaged setting. Writing the drag as $\tilde \tau = \tilde \rho \tilde
A g\cos(\theta) \mu$, we note from~\eqref{eq:steady balance} that $\mu =
\tan\theta$ for steady uniform flows.  Using this, we
substitute~\eqref{eq:hstop} into~\eqref{eq:U grains} and rearrange to find the
following friction law that preserves the section-averaged steady uniform
balance and is consistent with prior depth-averaged constructions:
\begin{equation}
    \mu = 
    %\mu_s + \frac{\mu_d - \mu_s}{1 + \frac{\xi}{\tilde{\mathcal{L}}}
    %\frac{\int_{-\infty}^{\infty}\tilde{h}^{5/2}\,\mathrm{d}\tilde{y}}{\frac{\tilde{A}\tilde{U}}{\sqrt{g\cos\theta}} + \gamma
    %\int_{-\infty}^{\infty}\tilde{h}^{3/2}\,\mathrm{d}\tilde{y}}},
    \mu_s + 
    (\mu_d - \mu_s) \left(
    1 + \frac{\xi}{\tilde{\mathcal{L}}}
    \frac{\int_{-\infty}^{\infty}\tilde{h}^{5/2}\,\mathrm{d}\tilde{y}}{\frac{\tilde{A}\tilde{U}}{\sqrt{g\cos\theta}} + \gamma
    \int_{-\infty}^{\infty}\tilde{h}^{3/2}\,\mathrm{d}\tilde{y}}
    \right)^{-1}\!\!,
%    =
%    \mu_s + \frac{\mu_d - \mu_s}{1 + \frac{\xi\tilde{s}_0}{\tilde{\mathcal{L}}}
%    \frac{\mathcal{J}(5/2)}{AU\Fr\mathcal{J}_0(1)^{3/2}/\mathcal{J}_0(0)
%    + \Gamma \mathcal{J}(3/2)}}
    \label{eq:granular friction}%
\end{equation}
where $\mu_s = \tan\theta_s$ and $\mu_d = \tan\theta_d$.
The corresponding non-dimensionalised 
granular drag law is given by
\begin{equation}
    \tau = A \mu / \mu_0.
    \label{eq:granular drag}%
\end{equation}

\subsubsection{Critical Froude number}
After non-dimensionalising~\eqref{eq:U grains}, differentiating and rearranging,
it may be determined that
\begin{equation}
    Q_0' = \frac{5}{2}\frac{\mathcal{J}_0(3/2)}{\mathcal{J}_0(5/2)}
    + \gamma \frac{\mathcal{J}_0(0)^{1/2}}{\Fr\mathcal{J}_0(1)^{3/2}}
    \left(
    \frac{5}{2}\frac{\mathcal{J}_0(3/2)^2}{\mathcal{J}_0(5/2)}
    - \frac{3}{2}\mathcal{J}_0(1/2)
    \right)\!.
    \label{eq:Q0p granular}%
\end{equation}
This generalises the formula of~(\ref{eq:Delta Q0p power law single}\emph{b})
with $p=3/2$, to the case of arbitrary $\gamma$.

\begin{figure}
    \begin{center}
        \includegraphics[width=\textwidth]{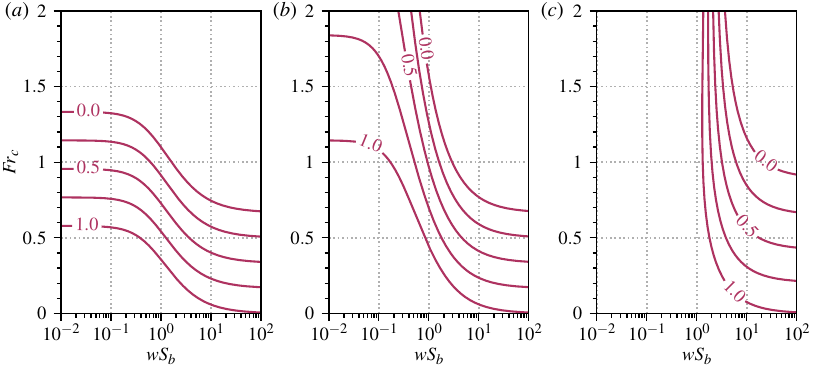}%
    \end{center}
    \caption{Critical Froude number curves for granular flow 
    in a trapezoidal channel. 
    Within each panel, separate curves show $\Fr_c$ for $\gamma = 0, 0.25, 0.5,
    0.75$ and $1$. In each case, flow is unstable for parameters above the
    corresponding curve.
    Panel~(\emph{a}) plots the case where shape factors are
    neglected, obtained via~\eqref{eq:Frc gran no shape}, while panels
    (\emph{b,c}) employ the shape factor function of~(\ref{eq:trap geom
    params}\emph{c}) with $p=3/2$ and (\emph{b}) $\chi_\infty = 1$, (\emph{c})
    $\chi_\infty = 5/4$.}
    \label{fig:gran Frc}%
\end{figure}

Since~\eqref{eq:Q0p granular} depends on the Froude number, when $Q_0' = U_0' -
\Delta$ is substituted into~\eqref{eq:sectional trowbridge} the resulting
expression must be rearranged to obtain $\Fr_c$.
When shape factors are neglected, this leads to
\begin{equation}
    \Fr_c = \frac{2 - \frac{\gamma}{[\mathcal{J}_0(0)\mathcal{J}_0(1)]^{1/2}}\left(
    5\frac{\mathcal{J}_0(3/2)^2}{\mathcal{J}_0(5/2)}
    - 3\mathcal{J}_0(1/2)
    \right)}{3 + 5\left(
    \frac{\mathcal{J}_0(1)\mathcal{J}_0(3/2)}{\mathcal{J}_0(0)\mathcal{J}_0(5/2)}
    - 1
    \right)}.
    \label{eq:Frc gran no shape}%
\end{equation}
This formula may be specialised for different channels studied
herein by computing $\mathcal{J}_0(n)$.
In the limit of an infinitely wide channel,
$\mathcal{J}_0(m)/\mathcal{J}_0(n) \to 1$, for any $m, n \geq 0$. Using this, it
may be verified that $\Fr_c \to 2(1-\gamma)/3$ in this limit, which is the
formula for the stability boundary of the unconfined
problem~\citep{Forterre2003}.
Some example curves are plotted in figure~\ref{fig:gran Frc}(\emph{a}), for the
trapezoidal geometry by evaluating~\eqref{eq:Frc gran no shape}
using~\eqref{eq:J(n) trap}.
As $\gamma$ increases from zero, $\Fr_c$ is lowered for all $wS_b$ values.
To a reasonable approximation, each curve is simply a translation in $\Fr_c$ of
the $\gamma = 0$ stability boundary.

The inclusion of shape factors complicates the calculation of $\Fr_c$. However,
for a given case, it may be obtained by using a numerical root finding algorithm
to isolate the critical cases of the~\eqref{eq:sectional trowbridge}
inequalities. We do this using~\eqref{eq:Q0p granular} and the~(\ref{eq:trap
geom params}\emph{c}) shape factor formula with $\chi_\infty = 1$ to produce the
curves in figure~\ref{fig:gran Frc}(\emph{b}).
Once again, each curve is close to a translation in $\Fr_c$ of the $\gamma = 0$ case.
However, while the $\gamma = 0$ curve tends to infinity as $wS_b\to 0$ (as
demonstrated in \S\ref{sec:trapezoids}), the $\gamma > 0$ curves do not feature
a singularity. Therefore, triangular channels feature an instability in
this case, and become more unstable as $\gamma$ increases.

The Bagnold scaling for steady unconfined flows of spherical grains implies  a velocity profile that rises
from the base in proportion with the depth to the power of $3/2$. 
From this, a shape factor of $\chi_\infty = 5/4$ may be
derived. \cite{Forterre2003} found that this increases the theoretical threshold
for instability of depth-averaged granular flows when $\gamma = 0$, but
questioned its usefulness, since $\chi_\infty = 1$ led to better agreement with
their experiments.  In figure~\ref{fig:gran Frc}(\emph{c}) we plot the case of
$\chi_\infty = 5/4$ and find similarly, that it greatly reduces the range of
unstable $\Fr$.  
When $\gamma = 0$, the $\Fr_c$ curve asymptotes to infinity at
$wS_b = 1 + 3\sqrt{5}/5$, which may be determined by solving
the~(\ref{eq:sectional trowbridge single}\emph{b}) inequality.  Like
figures~\ref{fig:gran Frc}(\emph{a,b}), as $\gamma$ increases, we see that its
effect is generally to reduce $\Fr_c$.
%,
%and we find that in the $wS_b\to\infty$ limit of unconfined flow, $\Fr_c \to
%\sqrt{(4+\gamma^2)/5}-\gamma$.
Additionally, though it is barely detectable within
the range of the figure, the $\gamma > 0$ curves briefly turn back on themselves
before reaching their vertical asymptote.  While this gives rise to values of $wS_b$
for which flows are theoretically unstable over a only a finite region of $\Fr$,
this effect is very slight and consequently, we do not explore it further. For
the physical problem, the dependence of the spatial growth rate on $\Fr$ and
$wS_b$, which we shall now address, is much more important.

\subsubsection{Growth rate and saturation amplitudes}
\begin{figure}
    \includegraphics[width=\columnwidth]{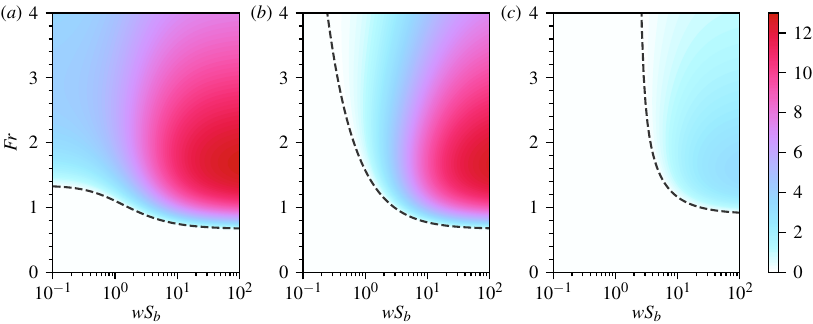}%
    \caption{%
    Maximum spatial growth rate for granular drag in a trapezoidal channel, with
    $\tilde L/\tilde D_0 = 1000$ and different shape factor choices:
    (\emph{a})~$\shape = 1$, (\emph{b,c})~$\shape$ set according
    to~(\ref{eq:trap geom params}\emph{c}) with (\emph{b})~$\shape_\infty = 1$
    and (\emph{c})~$\shape_\infty = 5/4$.  Dashed lines show the stability
    boundaries in each case, determined via~\eqref{eq:sectional trowbridge}.
    Note that within the stable regime, the maximum spatial growth is zero, due
    to the presence of a neutral mode at $k=0$ [see~\eqref{eq:growth rate formula}]. 
    The following illustrative parameters
    are used in the friction law [\eqref{eq:granular friction}]: $\mu_s =
    \tan(20^\circ)$, $\mu_d = \tan(34^\circ)$ and $\xi\tilde
    D_0/\tilde{\mathcal{L}} = 1.5$.
    }
    \label{fig:spatial growth granular}%
\end{figure}

The maximum spatial growth rate, 
measured with respect to the channel length $\tilde L$
was given in Eq.~\eqref{eq:asym spatial growth}. It may be
computed for the granular drag law of~\eqref{eq:granular friction}
and~\eqref{eq:granular drag}, by using the steady balance~\eqref{eq:steady balance}
to replace $\tan\theta$ with $\mu_0$,
in addition to
%
%\begin{equation}
%    \frac{\tilde L}{\tilde D_0} \frac{\real(\sigma)\mathcal{F}_0}{\Fr\left[\Fr + \sqrt{1 +
%    \Fr^2(\shapefu^2/4 + \shapefs)}\right]},
%    \label{eq:spatial growth granular}%
%\end{equation}
%
computing the relevant derivatives of $\tau$, in conjunction with the formulae
from Eqs.~\eqref{eq:trap geom params all} to determine
$\real(\sigma)$ and the shape factor terms.  The magnitude of the spatial growth
rate is
set by the ratio $\tilde L/\tilde D_0$.  Since the channel length is arbitrary,
we set $\tilde L/\tilde D_0 = 1000$ -- a reasonable order of magnitude for both
laboratory and field scale flows -- and plot a colour map of the spatial growth rate
in figure~\ref{fig:spatial growth granular}, using illustrative friction
parameters given in the figure caption.

The $\shape = 1$ case is shown in figure~\ref{fig:spatial growth
granular}(\emph{a}).  Maximal growth over all parameter space is attained in the
hydraulically wide regime, when $\Fr \approx 1.7$. In the low $wS_b$ regime,
spatial growth rates are typically no more than one third of this value.
Consider roll waves emerging from an upstream source perturbed by random noise.
Assuming that the characteristic distance between the source and the first
observations of roll waves is controlled principally by the linear stability
mechanism, this implies that waves in a triangular channel should appear roughly
three times further downstream than the equivalent unconfined analysis would
predict. In some cases, this means that waves will not be observed at all.  

Figure~\ref{fig:spatial growth granular}(\emph{b}) shows the growth rate when
the shape factor given in~(\ref{eq:trap geom params}\emph{c}) is
incorporated, but with the vertical shear neglected by setting $\shape_\infty =
1$. This substantially expands the stable
region of parameter space, compared with figure~\ref{fig:spatial growth
granular}(\emph{a}) and reduces growth rates in the low $wS_b$ regime.
Even more striking is the case $\shape_\infty = 5/4$, plotted in
figure~\ref{fig:spatial growth granular}(\emph{c}) which accounts for the
shear of the vertical velocity profile.
Here, the spatial growth rate
is even more severely restricted -- the maximum reaches only roughly $30\%$ of
the maxima in figures~\ref{fig:spatial growth granular}(\emph{a,b}) and a broader region
is linearly stable. 

In summary, figure~\ref{fig:spatial growth granular} encapsulates much of the
intuition gained from the analytical work up to this point. The geometry of a
convex open channel inhibits the existence of roll waves and reduces their
linear growth rates in unstable regions.  These effects are further exacerbated
by the presence of a cross-stream velocity profile induced by the channel shape,
which renders triangular channels unconditionally stable to roll waves.  We now
validate this intuition by reporting laboratory experiments of dry granular
flows and separately, numerical investigations of unsteady flow dynamics.

\section{Nonlinear dynamics}
\subsection{Granular experiments}
\label{sec:experiments}%
Experiments were conducted in $3\,\mathrm{m}$-long channels with three different
cross-sections: two trapezia with base widths $3.0\,\mathrm{cm}$ and
$1.5\,\mathrm{cm}$, and a triangle.  Each channel was constructed from strips of
$8\,\mathrm{mm}$-thick acrylic sheeting placed side-to-side and covered on one
face with a single length of skateboard grip tape, composed of glass beads
emplaced within a flexible polymer layer.  The banks are maintained at fixed
angles using 3D-printed supports.  The resulting channels were then mounted on a
supporting platform inclined at $32^\circ$ to the horizontal, as measured using
a digital inclinometer.  Within these channels, flows of monodisperse glass
beads, sieved to lie within the $200$--$300\,\mu\mathrm{m}$ diameter range, were
initiated from a funnel with an adjustable valve.  In comparison, the standard
deviation of the grip tape surface was measured to be $260\,\mu\mathrm{m}$,
which is sufficient to ensure no-slip at the base~\citep{Bougouin2026}.
Time-dependent measurements of  the mass flux $\tilde{Q}_m$ of grains exiting
the channel were obtained with a scale, alongside measurements of the flow
surface at $11$ equispaced downstream locations.  The latter data were recorded
with a scanCONTROL 2950-10 laser scanner from Micro-Epsilon.  The mean surface
elevation $\tilde{s}$ is then determined by averaging these measurements over
the cross-stream direction, while the local flow height $\tilde{h}$ is measured
relative to the bed elevation prior to the onset of flow. This is used to
compute the flow area $\tilde{A}$ and hydraulic depth $\tilde{D}$ via numerical
evaluations of~\eqref{eq:area} and~\eqref{eq:hydraulic depth}.

The experimental procedure was as follows. Using the funnel valve, flows were
initiated in the widest chute at many different fluxes, to find conditions where
roll waves were produced most readily downstream.
The Froude number was estimated for the flow 37\,cm from the inlet, upstream of
any visible surface waves, using the formula
\begin{equation}
    \Fr = \frac{\langle \tilde Q_m
    \rangle}{\tilde{\rho}_g \phi_g\langle\tilde
    A\rangle\sqrt{g\cos(\theta)\langle\tilde{D}\rangle}},
    \label{eq:Fr expt}%
\end{equation}
where $\tilde{\rho} = 2500\,\mathrm{kg\,m}^{-3}$ is the density of the grains,
$\phi_g = 0.53$ is a volume fraction that they are assumed to occupy during
the flow (chosen to match measurements from discrete element method simulations at
$\theta = 29^\circ$ conducted by
\cite{Gadal2026}), and angled brackets denote the time average of a given
quantity, taken over statistically steady flow conditions.  Given the findings
of \S\ref{sec:tilted trapezoids}, care was taken to minimise any lateral tilt
present in the channel by adjusting its supports, using both visual
observations of the flow and laser-line readings of the channel base to detect.
During this, the presence or absence of waves was noted to be very sensitive to
small variations in tilt, in line with the theoretical stability boundaries
plotted in figure~\ref{fig:tilt}.  The largest and fastest growing waves were
observed when $\Fr \approx 0.8$.  Following this, flows were initiated in the
two narrower channels, with the flux adjusted to maintain approximately the same
base Froude number.

\begin{figure}
    \includegraphics[width=\textwidth]{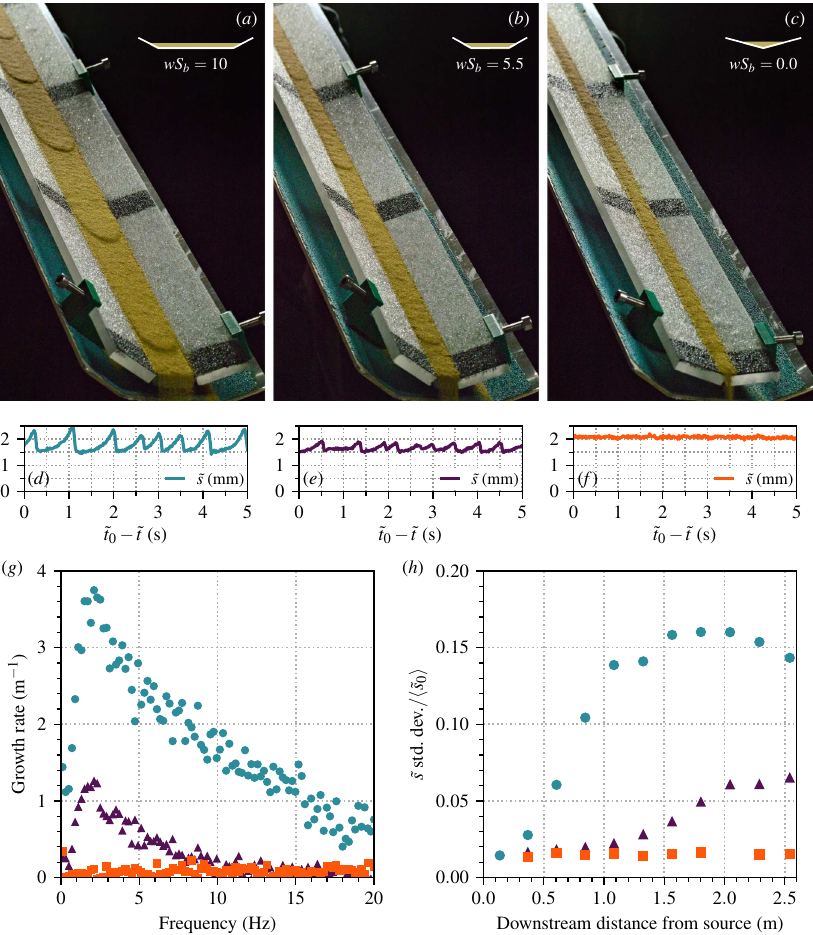}%
    \caption{Data from granular chute experiments
    in three different trapezoidal channels, 
    inclined at $\theta = 32^\circ$ to the horizontal,
    with
    (\emph{a})~$\tilde w = 3.0\,\mathrm{cm}$, $wS_b = 10$,
    (\emph{b})~$\tilde w = 1.5\,\mathrm{cm}$, $w S_b = 5.5$, (\emph{c})~$\tilde w
    = 0.0\,\mathrm{cm}$, $w S_b = 0.0$.
    %Flow propagates from from a hopper (out of
    %frame), which is laden with monodisperse glass beads ($200$--$300\mathrm{\mu
    %m}$ in diameter).  
    Steady elevation values were inferred using laser data measured 
    37\,cm from the flow inlet and used to determine
    $wS_b=\tilde{w}S_b/\langle\tilde{s}_0\rangle$ and $\Fr$,
    via~\eqref{eq:Fr expt}.
    In each case, the upstream flux was adjusted to give an initially steady
    flow at $\Fr = 0.85\pm 0.11$.
    The channel shape and wetted region in steady conditions are 
    shown in the top right at 1:1 aspect ratio.
    Measurements of surface elevation $\tilde{s}$, recorded 230\,cm from
    the inlet, are shown in (\emph{d--f}), directly beneath the corresponding
    experimental still. These panels show a small representative sample of the
    data recorded and are plotted as a function of time, relative to a reference
    time of $\tilde{t}_0 = 120\,\mathrm{s}$ into the experimental run.
    (\emph{g})~Linear spatial growth rates estimated for different temporal
    frequencies of the measured surface elevation signals.  (\emph{h})~Standard
    deviation of the surface elevation along the slope, normalised with respect
    to $\langle\tilde{s}_0\rangle$.
    }
    \label{fig:grain expt}%
\end{figure}

Data from the three experiments are shown in
figure~\ref{fig:grain expt}.
Firstly, figure~\ref{fig:grain expt}(\emph{a}) is a photograph of the lower half
of the $\tilde{w}=3.0\,\mathrm{cm}$ trapezoidal channel, during the flow. Four
coherent waves are clearly visualised from the shadows they cast, with a further
just noticeable in the top left of the frame.  In the corresponding
$\tilde{w}=1.5\,\mathrm{cm}$ trapezium image of figure~\ref{fig:grain
expt}(\emph{b}), the waves are harder to discern, though at least three are
present. Conversely, in figure~\ref{fig:grain expt}(\emph{c}), which shows the
$\tilde{w}=0.0\,\mathrm{cm}$ case, there are no observable waves.  Typical
measurements from the laser scanner are plotted for the $\tilde{w} = 3.0,1.5$
and $0.0\,\mathrm{cm}$ cases in figure~\ref{fig:grain expt}(\emph{d}--\emph{f})
respectively. These confirm the observation that the wave amplitudes decrease as
the channel narrows.  For the triangular chute, the only noticeable surface
disturbances are from measurement noise, some of which is likely attributable to
random kinetic granular collisions occurring at the top of the flow. On
adjusting the flux to several different values no waves were observed to form
spontaneously. In the case with greatest flux, $\Fr$ was measured to be $1.7$ at
84\,cm from the inlet. To assess whether waves can form at all in this geometry,
we generated large-amplitude disturbances by perturbing the flow directly by
tapping it with the tip of a paintbrush to create regions where grains were
unevenly distributed. The resulting perturbations decayed rapidly in all cases.
This procedure for the figure~\ref{fig:grain expt}(\emph{c}) experiment can be
viewed in Supplemental Movie~1.  It is also instructive to observe the decay of
perturbations at higher flux, to aid visualisation of the cross-stream flow
structure.  In Supplemental Movie~2, we disturb such a flow on a $26.5^\circ$
slope by emptying a small vessel containing $13\,\mathrm{g}$ of red glass beads
(in the same diameter range) onto the surface.  Instead of forming a steep shock
front, the initially thicker red region becomes rapidly and continually
stretched out, under the influence of the underlying cross-stream velocity
profile.  Consequently, the disturbance decays until the red beads are fully
absorbed into the steady base flow.

The remaining panels of figure~\ref{fig:grain expt} show data pertaining to the
spatial growth of the instability. The linear spatial growth rate is quantified in the
dispersion relation of figure~\ref{fig:grain expt}(\emph{g}).  This is obtained
by taking temporal Fourier transforms of the flow elevation data at each
recorded point downstream of the inlet, then fitting an exponential growth model
over the region of linear growth for each frequency.  The resulting data imply
that during the early stages of development, waves in the
$\tilde{w}=1.5\,\mathrm{cm}$ channel grow at roughly one third of the rate of
those in the wider channel.  The fastest growing frequencies in these cases are
between $1$--$3$\,Hz, which is commensurate with the signals observed downstream
in figures~\ref{fig:grain expt}(\emph{d,e}).  Conversely, for the triangular
geometry, no particular frequency dominates and the growth rates that are
computed reflect noise in the measurements, rather than the coordinated growth
of perturbations.  The nonlinear development of the wave amplitudes is captured
in figure~\ref{fig:grain expt}(\emph{h}). This shows the standard deviation of
surface elevation as a function of downstream distance, which may be used as a
proxy for the characteristic wave amplitude.  Using the data for the steady
triangular chute flow (orange squares) to quantify typical baseline fluctuations
of the free surface, the data for the widest chute begin to diverge from this at
approximately $0.5\,\mathrm{m}$ from the inlet, while in the
$\tilde{w}=1.5\,\mathrm{cm}$ case, this occurs at around $1.5\,\mathrm{m}$,
which is commensurate with the figure~\ref{fig:grain expt}(\emph{g}) dispersion
relation.  We further see that the $\tilde{w}=3.0\,\mathrm{cm}$ waves saturate,
reaching a maximum standard deviation of around $0.25\,\mathrm{mm}$, while the
corresponding data points for the $\tilde{w}=1.5\,\mathrm{cm}$ case reach less
than half this value.  Whether these latter waves are fully saturated is
unclear. However, they appear unlikely to reach the size of the waves in the
widest chute.

These experiments confirm the essential predictions of the linear theory: namely
that, as the base width of the trapezoidal channel narrows, the growth rates of
roll waves diminish and eventually become stable.
Moreover, they indicate that these trends in linear theory presage the nonlinear
dynamics. Specifically, waves in narrower channels arise further downstream and
attain lower amplitudes.

\subsection{Case study: simulations and travelling wave solutions for an Alpine
debris flow}
\label{sec:numerics}%
An alternative approach for investigating the roll wave instability in
non-rectangular channels is to use computational methods. This is particularly
useful for flows that are not directly experimentally accessible, or difficult
to measure. For example, although some particle-dense debris flows can be
successfully modelled using single-phase governing equations with a granular
friction law~\citep{Aaron2025}, their physics cannot be easily captured by
analogue experiments, since it does not readily rescale to a laboratory setting.
The presence of water in field-scale debris flows reduces their basal friction,
enabling them to flow on far shallower slopes than dry grains, while `wet'
experiments introduce the challenge of matching additional physical processes
such as the physics of water--grain interactions and the mixture rheology at
small scales.

Therefore, motivated by a recent proliferation of high-quality roll wave
observations in debris
flows~\citep{Schoffl2023,Chen2024,Aaron2025,Spielmann2025}, in this subsection
we report some numerical simulations in dimensional units at field scale, and use
them to examine the effect of channel geometry on these natural hazards,
alongside both the linear theory and nonlinear roll wave solutions, which we
construct using methods developed in Appendix~\ref{appendix:tw solns}.

Figure~\ref{fig:example channels} demonstrates a variety of different debris
flows channels in the Swiss Alps, from a broad, roughly trapezoidal section, to
an almost triangular channel.
\begin{figure}
    \includegraphics[width=\textwidth]{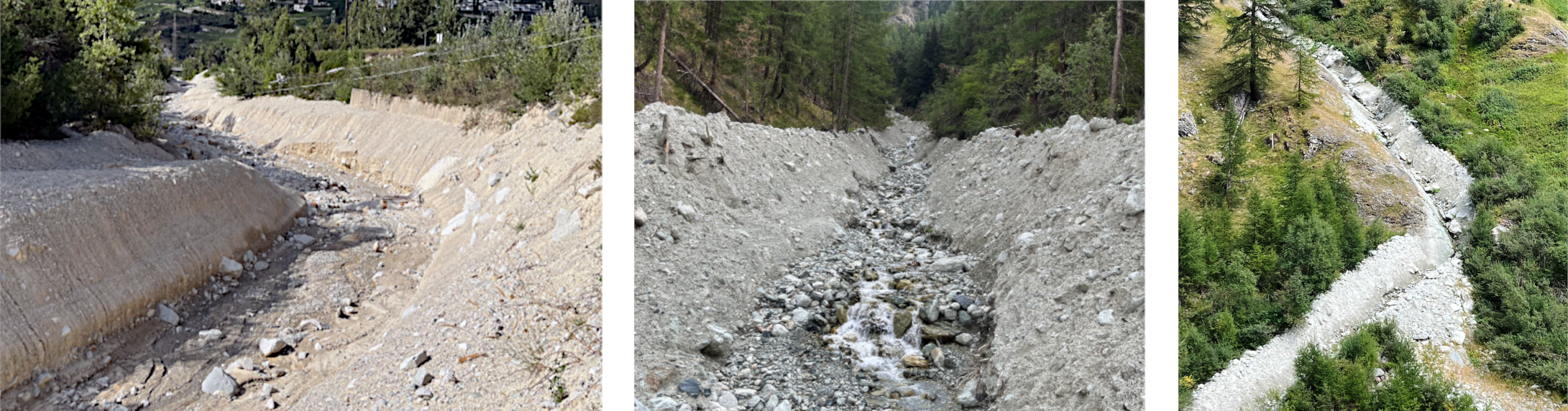}%
    \caption{Three natural channels in the Swiss Alps that are susceptible to
    rainfall-triggered debris flows. From left to right: Illgraben, Grand
    Torrent and Torrent de Veisivis. The widest of these (Illgraben) sustains
    multiple events each year that contain roll waves.
    }%
    \label{fig:example channels}%
\end{figure}
The widest of these is the continuously monitored Illgraben valley torrent,
which forms the starting point for our simulations.
We follow the study of~\cite{Aaron2025}, who modelled the
development of roll waves in this catchment
using~\eqref{eq:governing A 1}, \eqref{eq:governing A 2} and the granular
friction law of~\eqref{eq:granular friction} in the unconfined limit, where the
equations reduce to a one-dimensional depth-averaged model.
Removing this latter assumption, we simulate the generalised system with a
trapezoidal cross-section, but retain their frictional parameters, which led to
good quantitative agreement with measured field data.  These are: $\mu_s =
1.5\times 10^{-4}$, $\mu_d = 0.14$, $\xi = 1$, $\tilde{L} = 0.9\,\mathrm{m}$ and
$\gamma = 0.8$.  The channel at Illgraben possesses a near constant slope of
approximately $\theta = 4.5^\circ$ along its length and we use this value in all
simulations.  In~\cite{Aaron2025}, a phenomenological adjustment to the friction
law due to~\cite{Edwards2019}, was used, which is needed to model conditions
near arrest. To avoid unnecessary complications, we do not include it here.
%
%In addition to simulations of the section-averaged system, we
%consider the set of 2D depth-averaged governing equations given in~\eqref{eq:2D
%1}--\eqref{eq:2D 3} with the same friction law,
%setting all shape factors to unity.
%This enables the cross-stream dynamics of mass and momentum to be explicitly
%simulated and provides a nice comparison with the section-averaged theory.
%
%In both cases, 

Simulations are performed using the Basilisk software's finite volume solver for
the Saint-Venant equations~\citep{Popinet2013}, which implements a
central-upwind scheme due to \cite{Kurganov2002}.  This was adapted to
integrate~\eqref{eq:governing A 1} and~\eqref{eq:governing A 2} in conservative
form for a trapezoidal topography with the cross-stream shape factor given
in~(\ref{eq:trap geom params}\emph{c}) and $\chi_\infty = 1$, as well as the
granular friction law of~\eqref{eq:granular friction}. The grid spacing is
$0.15\,\mathrm{m}$ and a Courant-Friedrichs-Lewy condition of $0.5$ is used to
adaptively set the time step.  %\TODO{Share these modifications somehow.}

At the upstream boundary of each simulation, Dirichlet conditions are applied,
corresponding to a steady uniform flow state, whose wetted area is perturbed by
a low-amplitude random noise function $f$, which seeds the instability (if
present). Specifically, for the section-averaged
simulations, we set
\begin{subequations}
\begin{equation}
    \tilde{A} = \tilde{A}_0(1 + \epsilon f(\tilde{t})),\quad
    \tilde{U} = \tilde{U}_0,
    \tag{\theequation\emph{a,b}}%
    \label{eq:bcs both}%
\end{equation}
    \label{eq:bcs}%
\end{subequations}
with $\epsilon = 0.05$.  
%Then, for the 2D depth-averaged simulations, we set
%$\tilde{h} = \tilde{s}(\tilde{A}) - \tilde{b}$ at each upstream boundary point,
%using the same $\tilde{A}$ as~(\ref{eq:bcs}\emph{a}), alongside $\bar{u} =
%\bar{u}_0$, $\bar{v} = 0$ for the downstream and cross-stream velocities, with
%$\bar{u}_0$ determined by~\eqref{eq:ubar grains} with $\tilde{h} =
%\tilde{s}(\tilde{A}_0) - \tilde{b}$.  
For the noise function $f$, we generate a
uniform pseudorandom number in the interval $[-1,1]$ at each numerical time
step, using the C standard library. This choice is simple to implement and
produces results that statistically converge as the numerical space and time
resolutions are increased.
To select steady state values for~\eqref{eq:bcs both}, we fix a $\Fr$
and flux $\tilde{Q}_0 = \tilde{A}_0 \tilde{U}_0$, 
substitute these quantities into~\eqref{eq:U grains}, adapted for the
trapezoidal geometry, and solve numerically for $\tilde{A}_0$, given $wS_b$.
This then constrains $S_b$ and $\tilde{w}$, whose values may be deduced
after a little algebra.

\begin{figure}
    \includegraphics[width=\columnwidth]{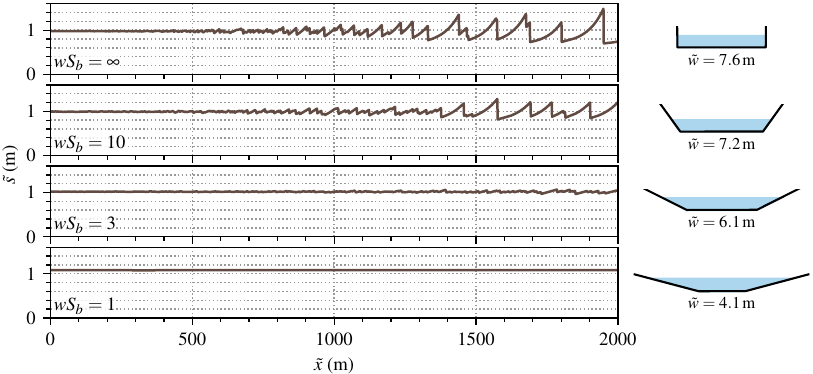}%
    \caption{
        Numerical simulations of the section-averaged system using parameters
        appropriate for debris flows. The Froude number and volume flux of the
        steady base flow are fixed at $\Fr = 0.6$, $\tilde{Q}_0 =
        14\,\mathrm{m}^3\mathrm{s}^{-1}$ via a Dirichlet boundary at $\tilde{x}
        = 0\,\mathrm{m}$, while the channel geometry varies between a rectangle
        with $wS_b = \infty$ and a trapezium with $wS_b = 1$ as indicated.  The
        simulation domain extends beyond the right-hand side of the figure and
        is longer than the region containing flow, thereby obviating for the
        need for a downstream boundary condition.  For reference, the diagrams
        on the right-hand side depict the channel geometry and the wetted area
        of the steady base flow in 1:1
        aspect ratio.
    }
    \label{fig:1dsim}%
\end{figure}

This procedure is used to construct figure~\ref{fig:1dsim}, which shows 
simulations with $\Fr = 0.6$, $\tilde{Q}_0 = 14\,\mathrm{m}^3\mathrm{s}^{-1}$
for four different channel geometries, with $wS_b = \infty,
10, 3, 1$ (from top to bottom).
%
%In the first case, a symmetry condition is applied at the boundaries to give
%frictionless sidewalls.
The topmost case is equivalent to the model used by~\cite{Aaron2025}, except
slightly simplified by setting constant mean volume flux and slope angle.  Small
perturbations at the left-hand boundary grow over $2\,\mathrm{km}$ to reach
amplitudes of up to $1\,\mathrm{m}$, which is commensurate with field
observations~\citep{Aaron2025}.  The geometry of the $wS_b = 10$ case is closer
to that of the actual Illgraben channel, whose lower reaches are banked and
around $7\,\mathrm{m}$ wide at the base~\citep{Aaron2025}. Roll waves are still
present at reasonable magnitudes, though they emerge farther down the channel
and at reduced amplitudes.  However, on narrowing the channel base to $\tilde{w}
= 6.1\,\mathrm{m}$ ($wS_b = 3$), the instability is barely observable.  Finally,
the bottommost case ($wS_b = 1$) lies beyond the stability boundary and contains
no roll waves.  This suggests that the prolific generation of surges in the
Illgraben is contingent on the particular geography of that site, rather than a
phenomenon that should be anticipated in all natural channels.  Furthermore, the
theoretical work of sections~\ref{sec:linear analysis} and~\ref{sec:examples}
leads us to believe that this implication is not qualitatively sensitive to the
details of the flow rheology.

\begin{figure}
    \includegraphics[width=\columnwidth]{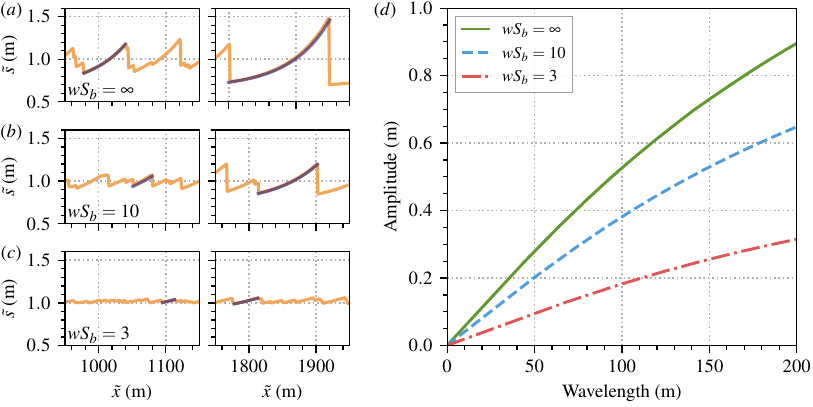}%
    \caption{%
        Nonlinear roll wave solutions for the debris flow case study.
        (\emph{a}--\emph{c}) Example solution profiles (purple) overlaid on
        snapshots of the numerical simulations presented in
        figure~\ref{fig:1dsim} (yellow). 
        (\emph{d})~Amplitude of surface elevation versus wavelength.
    }
    \label{fig:amplitudes}%
\end{figure}

To close this section, we address the amplitude of the nonlinear waves.  As
small roll waves grow, they merge with neighbouring waves, thereby acquiring
more volume and larger amplitudes.
In both the granular experiments and these simulations, channel confinement
reduced the amplitude of observed waves. However, it remains to check whether
this is purely because the diminished growth rate delays the onset of larger
waves, or if larger waves are dynamically inaccessible (i.e.\ they would never
be observed, even in an infinite channel).  The waves present in simulations,
which grow and coarsen in complicated ways, are transient manifestations of
simpler underlying periodic solutions to the governing equations.
These may be constructed for a given wavelength, following the approach of
\cite{Dressler1949}, which we generalise in Appendix~\ref{appendix:tw solns} for
arbitrary channels, rheologies and shape factors.
A selection of these travelling wave solutions are plotted in
figures~\ref{fig:amplitudes}(\emph{a}--\emph{c}), against snapshots of the
numerical simulations, demonstrating that they capture the time-dependent waves
well.
%
%We see that the solutions are indeed able to capture the properties of the
%waves in the time-dependent simulations. 
They may therefore be used to directly quantify the properties of roll waves in
different channels.

In figure~\ref{fig:amplitudes}(\emph{d}), we use this approach to plot
wave amplitude versus wavelength for the three unstable cases of
figure~\ref{fig:1dsim}.  We find that as waves grow to higher
amplitudes, their aspect ratio stretches.  This provides a mechanism for
constraining their ultimate size. As demonstrated by~\cite{Balmforth2004b} (for
hydraulically wide water flows) and~\cite{Razis2014} (for the granular case), roll
waves beyond a certain size are themselves vulnerable to a subharmonic
instability that causes them to split up, or otherwise shed mass.  The tension
between coarsening and splitting leads to a natural saturation amplitude, which
is system-dependent and unknown \emph{a priori}.  However, it is plausibly
related to the aspect ratio of the waves, since longer waves (for fixed
amplitude) more closely resemble the unstable uniform base flow.
Figure~\ref{fig:amplitudes} therefore suggests that narrower channels produce
smaller waves, even in the theoretical limit of arbitrarily long channels.

\section{Discussion}
\label{sec:discussion}%
To conclude, channel confinement mollifies roll waves in three ways: (1)~it can
quash instabilities that would be present in less confined settings; (2)~it
depresses the linear spatial growth rate of unstable flows, thereby delaying the
onset of waves; and (3)~it diminishes the ultimate amplitudes of mature roll
waves.
Far from being a second-order effect, the channel geometry is of comparable
importance to the Froude number of the base flow.
The backbone of these findings is the linear stability analysis of
sectionally-averaged free surface flows, with arbitrary basal resistance and
shape factor closures.  This allowed us to reach general conclusions that are
largely independent of the flowing material (some possible exceptions are noted
below) and straightforwardly adapted to different channel geometries.  When flow
in the channel is shallow, in the sense of the wetted extent being wider than it
is deep, explicit formulae were obtained for the critical Froude number in
trapezoidal and power law sections, which may be useful for assessing the
stability of natural or man-made channels.

As the geometry varies away from the wide rectangular conduits typically
considered in experimental and analytical work, the cross-stream velocity
profile of the steady base flow becomes nonuniform.  The inclusion of this
dependence in the momentum equation acts to further stabilise the theoretical
growth rates.  In particular, it renders many model flows unconditionally stable
in a triangular geometry. This accords with our separate experimental
investigations of water and granular flows. Neither medium appeared to support
stable roll waves in this geometry, regardless of the flux through the channel,
and in spite of attempts to directly induce wave generation via large manual
perturbations.  In the granular case, perturbations can be seen to stretch
rapidly under the influence of the cross-stream shear, decaying away soon after.
A video of this is provided in Supplemental Movie~2.

A corollary to these findings is that flat-bottomed channels (such as
trapezia) are acutely sensitive to cross-stream tilt, as predicted by the linear
theory in~\S\ref{sec:tilted trapezoids}. In the laboratory, it is possible to
stabilise roll waves by adjusting the flow base just a degree or two off the
horizontal.  This carries implications for assessing the stability of natural
and man-made channels alike, which can often be approximately flat towards the
centre, but are seldom perfectly level in the transverse direction.  Moreover,
it may point towards the effect of channel bends on the instability, which force
deviations in the wetted region as the flow superelevates towards the outer
banks.

Our study was motivated in part, by field observations of debris flows.  Though it
was known for many years that debris flows can feature trains of free-surface
waves~\citep{Blackwelder1928,Pierson1980}, which have been speculated by some authors
to emerge from roll wave instabilities~\citep[e.g.][]{Fraccarollo2000,Zanuttigh2007},
recent advances in monitoring are finally revealing the spatiotemporal
development of these surges~\citep{Aaron2025,Wetter2026}.  This has made it
possible to demonstrate the relevance of the linear roll wave instability in
this setting and begin to obtain good agreement between numerical models and field
data~\citep{Aaron2025}. From our analysis, we reason that it will be important
to incorporate the effect of channel geometry into the hazard prediction tools that
emerge from these new studies, either by running models on measured
topographies, or by utilising the section-averaged approach.
Furthermore, while multiple site-specific factors 
control debris flow surge development 
(e.g.\ downslope angle, sediment availability and antecedent rainfall), our
results suggest that the most prolific catchments, such as the Illgraben,
Switzerland~\citep{Aaron2025} and Jiangjia Ravine, China~\citep{Wei2025}, would produce
fewer and milder roll waves if the channels were narrower.

Despite the generality of our analysis, a handful of limitations present
intriguing challenges for future study.  
Averaging flow equations over the channel cross-section can limit their
predictive power. In unconfined analyses, authors have attempted to overcome
this either by retaining vertical
momentum~\citep[e.g.][]{Benjamin1957,Yih1963,Balmforth2004a,RuyerQuil2012,Depoilly2024} or introducing
corrections for any turbulent motions and shear in the velocity
profile~\citep{Kranenburg1992,Forterre2003,Richard2024}.
In confined channels, it may be fruitful to pursue
stability analyses that include cross-slope momentum components.
This would allow for the possibility that the flow develops cross-stream
vortices, either as part of the base flow, as studied by~\cite{Gadal2026}, or
due to competing modes of instability~\citep[such as
in][]{Borzsonyi2009,dOrtona2020,Pearse2026}.  However, such systems are
analytically challenging and unlikely to yield results that are as general and
expressive as those presented here.  The shape factor terms that we include to
account for shear appear to capture a real effect of stabilisation by the
cross-stream velocity profile. Nevertheless, little is understood regarding how
to model them outside of steady uniform flow conditions and naive
choices can lead them to cause unphysical solutions~\citep{Saingier2016}.

Next, not all flows are well
captured by the single-phase description used herein. For example, within
water--sediment mixtures, there may be significant transfers of momentum between
the two constituents, as well as variations in the rheology depending on their
relative fractions.  This necessitates a multi-phase description, which may
promote additional routes to instability that violate the general principles
given herein. 
Likewise, for fluids that possess a yield stress, two-layer models are required
to capture their internal 
decomposition into sheared and unyielded (plug-like) zones.
%Furthermore, while the $\gamma$
%parameter included in the granular friction law provided a phenomenological
%approximation of the heightened resistive forces present when there is a yield
%stress, the motion of true yield stress fluids is complicated by their internal
%decomposition into yielded and unyielded (plug-like) regions.  
For debris flows, both such properties are present~\citep{Parsons2001},
suggesting that their influence on the analysis should be investigated in future
studies. Nevertheless, single-phase models already fare
impressively well for some flows~\citep{Aaron2025}.

One way of incorporating a `yield-stress-like' effect in single-layer granular
flow models is through the $\gamma$ parameter included as part of the friction
law in~\eqref{eq:granular friction}.  This allows for uniform inclined layer
solutions with zero velocity, whenever the hydraulic depth is below a threshold
value for incipient motion.  However, the physics of grains close to, or at,
arrest, is more complicated than this. Experiments show that for some layer
depths, static and dynamic steady states coexist as part of a hysteresis loop
that can be captured using phenomenological adjustments to the friction
law~\citep{Edwards2019}. The hysteretic frictional response supports undamped
avalanching `erosion--deposition' waves that have the same shape as roll
waves that form due to linear instability~\citep{Borzsonyi2008,Takagi2011,Edwards2015}, but can arise via a
different mechanism. When flow is fed
upstream via a constant flux that is too low to produce a uniform flowing layer,
it forms a pile, whose grains can be held in place by static friction
coefficients that exceed the corresponding dynamic values. As the pile grows,
its sides steepen until the frictional balance is overcome and it fails
catastrophically.  The resulting pulse thins out into a travelling wave separated
by deposit layers fore and aft~\citep{Borzsonyi2008,Takagi2011,Rocha2019}.
Because these waves coexist with stable static flows, they can also be triggered
by finite-amplitude perturbations to a deposit layer~\citep{Edwards2021}. Both
these destabilising mechanisms are likely to be less sensitive to channel shape,
since the resulting waves can propagate over channel-filling deposits that
obscure the underlying bed geometry. 

\begin{figure}
    \begin{center}
        \includegraphics[width=0.7\columnwidth]{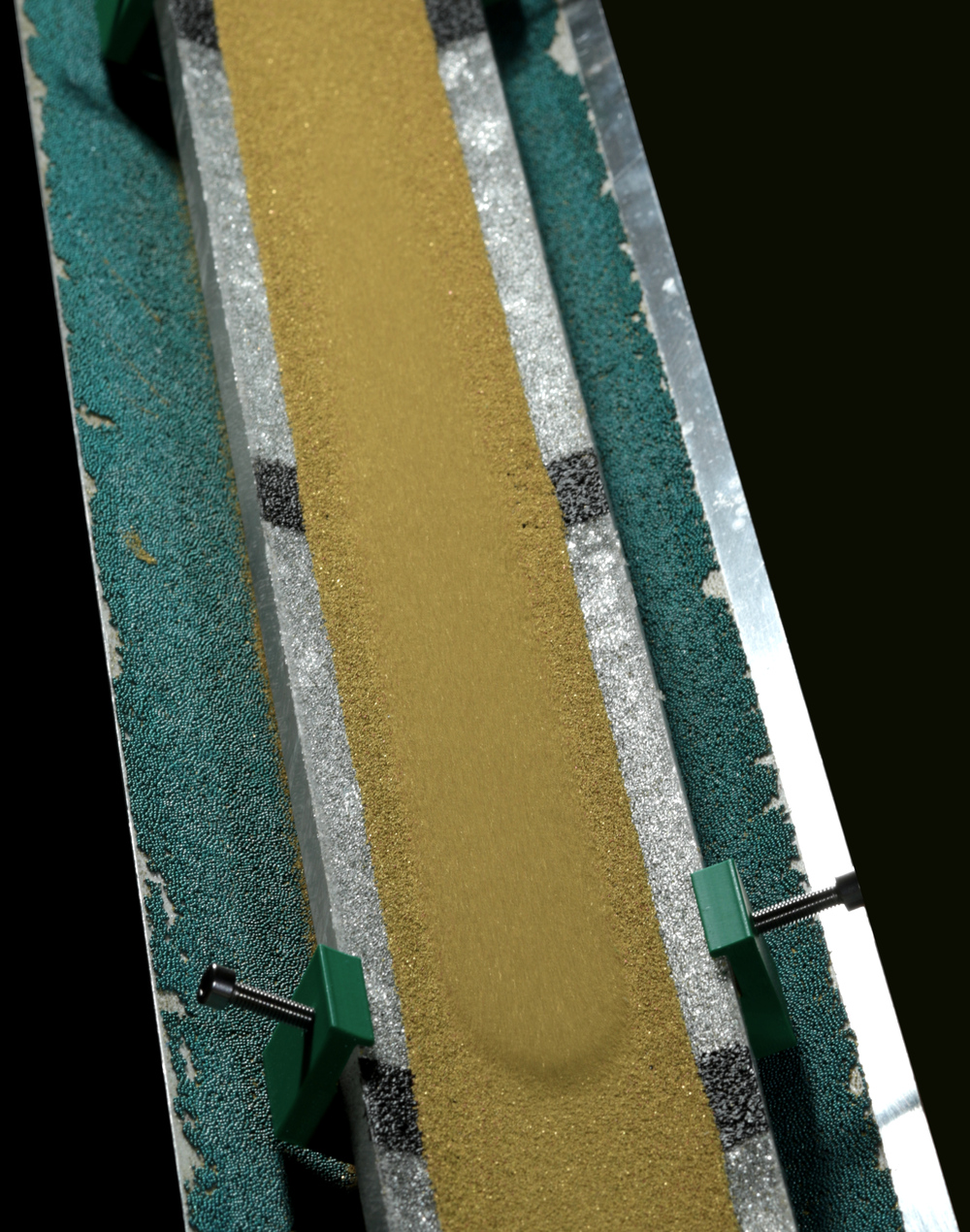}%
    \end{center}
    \caption{%
        Granular erosion--deposition wave in a triangular channel, inclined at
        $26.5^\circ$.
        The flowing region is distinguished by the presence of
        motion blur in the image.
    }
    \label{fig:ed wave}%
\end{figure}

Consequently, we find that it is indeed possible to produce stable
erosion--deposition waves in this way even in the triangular channel.  This is
demonstrated in Supplemental Movie~3, which was filmed after the upstream
material from experiment in Movie~2 drained, leaving a static deposit in
the channel.  The same perturbation that was rapidly absorbed by the steady
flowing layer of Movie~2 is applied once again.  On the deposit, it
develops into an undamped avalanche. The static grains are apparently less able than
the steady flow, to dissipate the energy of the disturbance, which can
continually recruit gravitational potential energy from the bed
by mobilising particles at its front.
A still image of such a wave (from a separate experiment) is shown in
figure~\ref{fig:ed wave}.
The flowing region, which forms an elongated teardrop shape, is just discernible from
the motion blur in the photograph.  
Note that it does not touch the sides of the deposit, indicating that the mobile
region is not in direct contact with the base.
Whether an analogous wave-forming mechanism
permits yield stress fluids (such as clay suspensions) or field-scale debris
flows to produce surges in channels that are linearly stable when flowing
remains to be seen.

\vskip 8pt
{\small{\setlength\parindent{0pt}\fontfamily{LinuxLibertineT-TLF}\selectfont\fontseries{sb}\selectfont
Acknowledgements.}
We thank Martin Quinn for assistance in constructing the experimental apparatus.
J.M.N.T.G.\ acknowledges Eric Bardou and Bob de Graffenried for organising the
field trip during which the figure~\ref{fig:example channels} photographs were taken.
Additionally, we acknowledge the use of multiple free software projects, most
notably Basilisk (https://basilisk.fr), as well as various numerical Julia and
Python libraries, including %(but not limited to)
DifferentialEquations.jl,
%~\citep{Rackauckas2017}, 
Roots.jl,
%~\citep{Verzani2020}, 
Interpolations.jl,
PerceptualColourMaps.jl,
Matplotlib,
%~\citep{Hunter2007},
NumPy,
%~\citep{Harris2020}, 
SciPy
%~\citep{Virtanen2020}
and Uncertainties (https://pythonhosted.org/uncertainties).
}
\vskip 8pt

{\small{\setlength\parindent{0pt}\fontfamily{LinuxLibertineT-TLF}\selectfont\fontseries{sb}\selectfont
Funding.}
This research was principally supported by funding from National Environment Research
Council (NERC) grants NE/X00029X/1 and NE/X013936/1.
C.G.\ acknowledges funding from the Royal Society through a Newton International
Fellowship (NIF/R1/231983).}
\vskip 8pt

{\small{\setlength\parindent{0pt}\fontfamily{LinuxLibertineT-TLF}\selectfont\fontseries{sb}\selectfont
Declaration of interests.}
The authors report no conflict of interest.}
\vskip 8pt

{\small{\setlength\parindent{0pt}\fontfamily{LinuxLibertineT-TLF}\selectfont\fontseries{sb}\selectfont
Data availability statement.}
The experimental data presented in \S\ref{sec:experiments}
are available in an online repository at \href{https://doi.org/10.5281/zenodo.22815714}{doi.org/10.5281/zenodo.22815714}.}

\appendix%
\section{Energy stability}
\label{appendix:energy stability}%
When $\shape \equiv 1$, physical intuition for the roll wave instability in a
general channel can be
obtained by deriving an energy equation.  Similar reasoning was previously
followed by~\cite{Trowbridge1987} for unconfined flows. By multiplying the
dimensionless mass
conservation relation
$\partial A/\partial t + (\partial/\partial x)(Au)=0$ by $\Delta As/\Fr^2$ and
the momentum equation~\eqref{eq:governing nondim 2} by $AU$, combining and
simplifying, we obtain
\begin{equation}
    \frac{\partial E}{\partial t}
    + \frac{\partial~}{\partial x}
    \left[
        \frac{1}{2}AU^3 + \frac{\Delta}{\Fr^2}AsU
        \right]
    = AU - \tau U,
    \label{eq:energy equation}%
\end{equation}
where $E$ 
is the sum of kinetic and gravitational potential energy within the channel
section, which we define by
denoting $h = \tilde{h}/\tilde{s}_0$, $y = \tilde{s}_0 \tilde{y} / \tilde{A}_0$,
and writing
\begin{equation}
    E = \frac{1}{2}AU^2 + \frac{\Delta}{2\Fr^2}
    \int_{-\infty}^{\infty}h^2 + 2hb\,\diff y.
\end{equation}
Substituting any of the modes in~\eqref{eq:modes} into~\eqref{eq:energy
equation} and averaging over a single wavelength gives
\begin{equation}
    \frac{\partial\langle E_1 \rangle}{\partial t}
    = \langle A_1 U_1 \rangle - 
    \frac{\partial \tau}{\partial s_0}\langle s_1 U_1 \rangle
    - \frac{\partial \tau}{\partial U_0}\langle U_1^2 \rangle
    \label{eq:energy pert}%
\end{equation}
where $\langle \cdot \rangle \equiv \frac{k}{2\pi}\int_0^{2\pi/k}\cdot\,\diff x$
and $E_1$, $A_1$ respectively denote the perturbation amplitudes of the energy
and cross-sectional area away from the base state.  The right-hand side terms
of~\eqref{eq:energy pert} are the power input to the disturbance by
gravitational forcing and the corresponding rate of loss via dissipation by the
basal stresses.  The amplitudes of the most critical modes, which occur at high
wavenumber, are easily computed as eigenvectors of the Jacobian matrix, giving
$s_1 = \pm \Fr$ and $U_1 = \Delta$ (up to normalisation).
%%
%\begin{equation}
%    \begin{pmatrix}
%        s_1 \\ U_1
%    \end{pmatrix}
%    =
%    \begin{pmatrix}
%        \pm Fr \\ \Delta
%    \end{pmatrix}.
%\end{equation}
%%
Moreover, $A_1 = A_0' s_1$ and it may also be determined that $A_0' = \Delta$
(with primes denoting differentiation with respect to $s$).
Therefore, the condition for the right-hand side to be positive at high $k$ is
\begin{equation}
    \pm \Fr \left(1 - \frac{1}{\Delta}\frac{\partial \tau}{\partial s_0}\right)
    > \frac{\partial \tau}{\partial U_0}.
\end{equation}
Given~\eqref{eq:steady balance diff} this is equivalent to the general stability
criterion of~\eqref{eq:sectional trowbridge simple}.  In other words, the
condition for linear instability (when $\chi \equiv 1$) is equivalent to the
requirement that asymptotically small wavelength disturbances gain energy from
the gravitational field at a rate quicker than can be dissipated via frictional
losses.

\section{Width-and-depth-averaged governing equations}
\label{appendix:width averaging}%
Here, we connect the analysis of laterally shallow channel flows in
the main text (\S\ref{sec:shallow flows}), to two-dimensional models that are commonly used
to study Earth-surface flows.
We assume that the full governing equations for the flow 
%(in Cartesian coordinates) 
may be averaged in the
$\tilde{z}$-direction, which we define to be perpendicular to the mean slope of the channel.
Denoting the downslope direction by $\tilde{x}$,
the cross-slope direction by $\tilde{y}$ and the corresponding velocity
components by $\tilde{u}$ and $\tilde{v}$, the depth-averaged velocity fields
$\bar{\vect{u}} = (\bar{u},\bar{v})$
are
\begin{equation}
    \bar{u} = \frac{1}{\tilde h}\int_{\tilde b}^{\tilde s} \tilde u
    \,\mathrm{d}\tilde z, \quad\mathrm{and}\quad
    \bar{v} = \frac{1}{\tilde h}\int_{\tilde b}^{\tilde s} \tilde v
    \,\mathrm{d}\tilde z.
\end{equation}
As in the main text, we assume the channel bed $\tilde{b}$ to be invariant in
the downslope direction.
The system for the flow height and depth-averaged
velocities may then be 
%derived by standard methods and 
given as
\begin{subequations}
\begin{gather}
    \frac{\partial \tilde h}{\partial \tilde t}
    + \frac{\partial~}{\partial \tilde x}(\tilde h \bar{u})
    + \frac{\partial~}{\partial \tilde y}(\tilde h \bar{v})
    = 0,\label{eq:2D 1}\\
    \frac{\partial~}{\partial \tilde t}
    (\tilde h \bar{u})
    + \frac{\partial~}{\partial \tilde x}
    \left(\chi_{uu}\tilde h \bar{u}^2
    + \frac{1}{2}g\cos(\theta) \tilde{h}^2
    \right)
    + \frac{\partial~}{\partial \tilde y}
    (\chi_{uv}\tilde h \bar{u}\bar{v})
    = g \sin(\theta) \tilde h - \frac{\tilde
    \tau_{\mathit{2D}}\bar{u}}{\tilde{\rho}|\bar{\vect{u}}|},\label{eq:2D 2}\\
    \frac{\partial~}{\partial \tilde t}
    (\tilde h \bar{v})
    + \frac{\partial~}{\partial \tilde x}
    \left(\chi_{uv}\tilde h \bar{u}\bar{v}
    \right)
    + \frac{\partial~}{\partial \tilde y}
    \left(
    \chi_{vv}\tilde h \bar{v}^2 +
    \frac{1}{2}g\cos(\theta)\tilde h^{2}
    \right)
    =
    -g \frac{\partial \tilde b}{\partial \tilde y} \tilde h
    - \frac{\tilde
    \tau_{\mathit{2D}}\bar{v}}{\tilde{\rho}|\bar{\vect{u}}|},
    \label{eq:2D 3}
\end{gather}
    \label{eq:2D all}%
\end{subequations}
where $\tilde{\tau}_{\mathit{2D}}$ is the basal drag acting on the flow and the
$\chi_{uu}$, $\chi_{uv}$, $\chi_{vv}$ terms are shape factors. The middle of
these is given by
\begin{equation}
    \chi_{uv} = 1 + \frac{1}{\tilde h}\int_{\tilde b}^{\tilde s} \left(
    \frac{\tilde{u}}{\bar{u}} - 1\right)
    \left(
    \frac{\tilde{v}}{\bar{v}} - 1\right)
    \,\mathrm{d}\tilde{z},~
%    \chi_{uu} = 1 + \frac{1}{\tilde h}\int_{\tilde b}^{\tilde s} \left(
%    \frac{\tilde{u}}{\bar{u}} - 1
%    \right)^2\mathrm{d}\tilde{z},~~
%    \chi_{vv} = 1 + \frac{1}{\tilde h}\int_{\tilde b}^{\tilde s} \left(
%    \frac{\tilde{v}}{\bar{v}} - 1
%    \right)^2\mathrm{d}\tilde{z},
\end{equation}
and the remaining two are defined analogously by replacing either of the velocity
fields with the other. 
%Note that $\chi_{uu} \equiv \chi_\infty$
%(see~\eqref{eq:shallow shape factors}).
Steady uniform flows occur when the cross stream velocity $\bar{v}_0$ vanishes:
\begin{subequations}
\begin{equation}
    \tilde{\tau}_{\mathit{2D}}(h_0, \bar{u}_0) = \rho g h_0 \sin(\theta),
    \quad
    \bar{v}_0 = 0.%
    \tag{\theequation\emph{a,b}}%
\end{equation}%
    \label{eq:steady 2D}%
\end{subequations}%

The section-averaged equations from the main text~(\ref{eq:governing
all}\emph{a,b}) may be obtained by integrating~\eqref{eq:2D 1}
and~\eqref{eq:2D 2} in the cross-stream direction. This causes the $\tilde{y}$
derivatives to vanish and~\eqref{eq:governing A 1} is recovered immediately
using~\eqref{eq:flux}.
Three terms in the derivation of~\eqref{eq:governing A 2} require brief
discussion.
Firstly, the cross-stream integral of the advective term implies that the
section-averaged shape
factor
\begin{equation}
    \chi = \frac{1}{\tilde A \tilde U^2}\int_{-\infty}^{\infty} \chi_{uu} \tilde h
    \bar{u}^2\,\mathrm{d}\tilde{y}.
\end{equation}
In~\eqref{eq:shallow shape factors}, this is approximated using the steady
uniform flow function $\bar{u}_0$ defined implicitly by~(\ref{eq:steady
2D}\emph{a}).
When this approximation is made, $\chi_{uu}$ is relabelled as $\chi_\infty$. For
the hydrostatic pressure term in~\eqref{eq:2D 2}, we note
that
\begin{equation}
    \frac{1}{2}\int_{-\infty}^{\infty}\frac{\partial~}{\partial \tilde x}
    \tilde{h^2}\,\mathrm{d}\tilde y
    = \tilde A \frac{\partial \tilde s}{\partial \tilde x}
\end{equation}
because $\partial \tilde h/\partial \tilde x = \partial \tilde s / \partial
\tilde x$.
Finally, integrating the downstream drag term implies that the aggregate
downslope drag is
\begin{equation}
    \tilde \tau = \int_{-\infty}^{\infty} 
    \frac{\tilde{\tau}_{\mathit{2D}}\bar{u}}{|\bar{\vect{u}}|}\,\mathrm{d}\tilde y.
\end{equation}
This expression approximates the integral over the wetted perimeter given
in~\eqref{eq:tau integral} for the case of a hydraulically wide channel.

\section{Channels with a power law dependence}
\label{appendix:power law}%
Consider flow with surface elevation $\tilde s_0$, in a channel whose
cross-section is given by a power law with exponent $m$.  In this Appendix, we
ask: what is the effect on the linear stability of varying $m$?  Therefore, we
define the channel base by the function $\tilde b(\tilde y) = \tilde s_0 |\tilde
y / \tilde y_0|^m$, with $\tilde y_0$ being half the top width of the flow
surface.  The case $m = 1$, of a triangular, or `V-shaped' cross-section, is
covered in the main text, as a special case of trapezoidal cross sections.  
As in~\S\ref{sec:shallow flows},
a shallow flow is assumed, with steady
depth-averaged downstream velocity given by a function $\bar{u}_0(\tilde h) =
\tilde K \tilde h^p$.

This reduces the problem of determining the critical Froude number to computing
the functional $\mathcal{J}(n)$, defined in~\eqref{eq:J(n)} for different values
of $n$, using this in~\eqref{eq:Delta Q0p power law} and \eqref{eq:chi power law}
to find $U_0' = Q_0 - \Delta$ and $\chi$, then using these in turn
in~\eqref{eq:sectional trowbridge}.
For the power law channel, $\mathcal{J}(n)$ is given by
\begin{equation}
    \mathcal{J}(n) 
    = \frac{2}{\tilde{s}_0^{n+1}}\int_0^{\tilde w s^{1/m}} 
    \left(\tilde s - \tilde s_0 (\tilde y / \tilde y_0)^m
    \right)^n\,\diff \tilde y.
%    \equiv \int_{-\infty}^{\infty} \tilde h^n\,\diff \tilde y
%    = 2\int_0^{(\tilde s/\alpha)^{1/m}} (\tilde s - \alpha \tilde
%    y^m)^n\,\diff \tilde y.
\end{equation}
By making the substitution
$Y = 1 - (\tilde y/\tilde w)^m s^{-1}$, the integrand simplifies to give
\begin{equation}
    \mathcal{J}(n) = 
    2\frac{s^{1/m+n}}{m}\frac{\tilde y_0}{\tilde s_0}
    \int_0^1 Y^n (1-Y)^{1/m-1}\,\diff Y = 
    %2 \frac{\tilde{s}^{1/m+n}}{m\alpha^{1/m}} 
    2\frac{s^{1/m+n}}{m}\frac{\tilde y_0}{\tilde s_0}
    B(n+1, 1/m),
    \label{eq:J(n) power law}%
\end{equation}
where $B$ is the beta function. Using the identity
$B(n,m)\equiv\Gamma(m)\Gamma(n)/\Gamma(m+n)$, where $\Gamma$ is the gamma
function, this may be evaluated
explicitly for some special cases. For example, $B(2,1/m) = m^2/(m+1)$ and
$B(n,1) = B(1,n) = 1/n$.
Furthermore, the gamma function identity $z\Gamma(z) \equiv \Gamma(z+1)$ implies
that $B(n+1, 1/m) = n(n+1/m)^{-1}B(n,1/m)$.

%The area $\tilde A$ is such a special case, where we note that
%%
%\begin{equation}
%    \tilde A = \mathcal{J}(1) = 2\frac{\tilde{s}^{1/m+1}}{m\alpha^{1/m}} B(2, 1/m)
%    = 2\frac{\tilde{s}^{1/m+1}}{\alpha^{1/m}} \frac{m}{m+1}.
%\end{equation}
Making use of these properties, by~\eqref{eq:Delta Q0p power law single} and~\eqref{eq:J(n)
power law}, we compute 
%$\Delta$ to be simply
%$\Delta = \tilde s_0/\tilde D_0 = \tilde s_0 \tilde A_0'/\tilde A_0$ to be simply
%
\begin{equation}
    \Delta = \frac{m+1}{m}\quad
    \mathrm{and}\quad
    Q_0' = \frac{1}{m} + p + 1.
\end{equation}
%
%For the computation of $U_0' = Q_0'-\Delta$ in Eq.~\eqref{eq:sectional
%trowbridge}, we first determine $Q = \tilde
%Q/\tilde Q_0$. 
%From the definition in Eq.~\eqref{eq:integrated power law}, this
%is
%%
%\begin{equation}
%    Q = \frac{\mathcal{J}(p+1)}{\mathcal{J}_0(p+1)} = s^{\frac{1}{m}+p+1},
%\end{equation}
%
Hence, $U_0' = p$ and by~\eqref{eq:sectional trowbridge simple}, if shape
factors are neglected
\begin{equation}
    \Fr_c = \frac{1 + 1/m}{p}.
    \label{eq:Frc power law no shape}%
\end{equation}

Then, for the shape factor,~\eqref{eq:chi power law} is used to give
\begin{equation}
    \shape
    = \shape_\infty \frac{B(2,1/m)B(2p+2,1/m)}{B(p+2,1/m)^2},
    \label{eq:beta power law}%
\end{equation}
where, as in the main text, $\shape_\infty$ is the shape factor for unconfined
flow.
The last expression in Eq.~\eqref{eq:beta power law}
is the most convenient form for numerical evaluation. However, switching to gamma
functions clarifies that
\begin{equation}
    \shape = \shape_\infty
    \frac{\Gamma(2)\Gamma(2p+2)\Gamma(p+2+1/m)^2}{\Gamma(2+1/m)\Gamma(2p+2+1/m)\Gamma(p+2)^2}\to
    \shape_\infty,\quad\mathrm{as}~m\to\infty.
\end{equation}
That is, as
the channel approaches a rectangle, the cross-stream dependence of the shape
factor becomes negligible.
Equation~\eqref{eq:beta power law} may be used to obtain exact expressions for
the cross-stream part of the shape factor $\chi/\chi_\infty$ for triangular
($m=1$) and parabolic ($m=2$) channels. 
Example values are given in
table~\ref{tab:shape factors}.
\begin{table}
  \begin{center}
\def~{\hphantom{0}}
  \begin{tabular}{ccccc}
      %\vspace{0.1cm}
      \diagbox{$m$}{$p$} & $1/2$ & $1$ & $3/2$ & $2$ \\
      1 & $\frac{25}{24}\approx 1.04$ & $\frac{9}{8}=1.125$ &
      $\frac{49}{40}=1.225$ & $\frac{4}{3}\approx 1.33$ \\[1.0ex]
      2 & $\frac{4096}{405\pi^2}\approx 1.02$ & $\frac{15}{14}=1.07$ &
      $\frac{262144}{23625\pi^2}\approx 1.12$ & $\frac{350}{297}\approx 1.18$
  \end{tabular}
      \caption{Cross-stream shape factors $\chi/\chi_\infty$ for triangular
      ($m=1$) and
      parabolic ($m=2$) channels.}
  \label{tab:shape factors}
  \end{center}
\end{table}
As may be expected, we see that $\chi/\chi_\infty$ is largest when the
depth-averaged velocity depends most acutely on flow depth (large $p$) and when
the channel base is far from a rectangle (small $m$).

\begin{figure}
    \begin{center}
        \includegraphics[width=\textwidth]{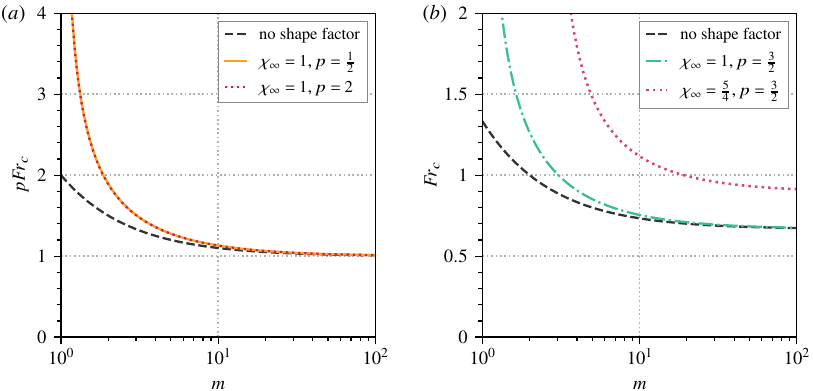}%
    \end{center}
    \caption{%
    Neutral stability curves for channels defined by a power law with exponent
    $m$.  (\emph{a})~Ratio of critical Froude number $\Fr_c(m)$ to its value in
    the limit $m\to\infty$ ($\Fr_c = 1/p$) for the case where shape factors are
    neglected, alongside cases with shape factors given by~\eqref{eq:beta power
    law} and $\chi_\infty = 1$, for $p = 1/2$, $p = 2$.  (\emph{b})~$\Fr_c$ for
    dry granular flow, $p=3/2$ (and $\gamma = 0$), with different shape factor
    choices.
    }%
    \label{fig:Frc power law}%
\end{figure}

From~\eqref{eq:sectional trowbridge}, and noting 
that $\shape$ carries no dependence on $s$ or $U$ (so $\shapefu =
2(\shape - 1)$ and $\shapefs = \shape - 1$),
the critical Froude number as a function of
$m$ is given in general, by
\begin{equation}
    \Fr_c = \left|
    \left(\frac{pm}{m+1} - (\shape-1)\right)^2
    -\shape(\shape-1)
    \right|^{-1/2}\!\!,
    ~~
    \mathrm{for}
    ~~
    \chi < 1 + \frac{p^2m^2}{(m+1)(m(2p+1) + 1)},
\end{equation}
with $\shape\equiv\shape(m)$ determined by Eq.~\eqref{eq:beta power law}.  As in
the case of the trapezoidal channel, the neutral curves collapse under rescaling
by $p$. When $\chi = 1$, this is exact, by~\eqref{eq:Frc power law no shape}.
Figure~\ref{fig:Frc power law}(\emph{a}) plots $p\Fr_c$ for this case, as well
as for when $\chi$ is set by~\eqref{eq:beta power law} with $\chi_\infty = 1$.
The $p = 1/2$ and $p=2$ curves are shown, which collapse near perfectly.
These curves may be compared to the corresponding curves for the trapezoidal
channel in figure~\ref{fig:Frcrit}, demonstrating that the exponent $m$ plays a
similar role to $wS_b$.  Finally, for completeness, in figure~\ref{fig:Frc power
law}{(\emph{b})}, the $\Fr_c$ curves for the granular ($p=3/2$, $\gamma = 0$)
case are shown, with the case $\chi_\infty = 5/4$ included. We note that,
similar to trapezoidal case (see figure~\ref{fig:gran Frc}), accounting for the
vertical dependence of the velocity profile in this way causes the unstable
region to retreat significantly towards higher $\Fr$.

Though there is scant experimental literature to confirm these conclusions, many
observations of granular flows in a 2\,m-long parabolic channel were conducted
by the authors during the study of~\cite{Gadal2026}. Using glass beads sieved in
$125$--$165\,\mu\mathrm{m}$ range, 21 steady flows with 
$\Fr \in [0.3, 8]$ were generated, and no roll waves were
recorded. This suggests that parabolic channels are indeed stabilising, since
the same granular material readily forms roll waves in a rectangular chute.

\section{Constructing nonlinear roll wave solutions}
\label{appendix:tw solns}%
Beyond linear order, it is possible to construct periodic solutions of the
governing equations that correspond to idealised versions of the waves
observed in unstable conditions. The method for doing so was originally 
established by~\cite{Dressler1949}
for unconfined flows. This is extended below to encompass the section-averaged
perspective. While the essential procedure is the same, it can be used to inform
us of the effect of confinement on the shape and size of roll
waves.

For this analysis, it is more convenient to work with the fields $A$, $U$ and
the equations in conservative form.  These may be obtained by defining $\tilde I
= \frac{1}{2}\int_{-\infty}^{\infty}\tilde h^2 \,\diff \tilde y$, noting that
$\partial \tilde I/\partial \tilde x = \tilde{A}\partial
\tilde{s}/\partial\tilde{x}$ and using this to rewrite the third term
of~\eqref{eq:governing A 2}. After non-dimensionalising the result along with~\eqref{eq:governing A 1}, one obtains
\begin{gather}
    \frac{\partial A}{\partial t} + \frac{\partial~}{\partial x}(AU) =
    0,\label{eq:gov conservative 1}\\
    \frac{\partial~}{\partial t}(AU)
    + \frac{\partial~}{\partial x}\left(\shape AU^2 + \frac{\Delta
    I}{\Fr^2}\right) = A - \tau,\label{eq:gov conservative 2}
\end{gather}
where $I \equiv \tilde I / (\tilde s_0 \tilde A_0)$.
A travelling wave solution
to these equations is
time-invariant in a reference frame that moves downstream with the velocity of
the wave.
Therefore, we recast the equations by transforming to a co-moving spatial
variable
$\eta \equiv x - c_wt$, where $c_w$ is a
constant wave speed to be determined. 
Under steady conditions, the resulting equations are
\begin{gather}
    \frac{\diff ~}{\diff \eta}[A(U-c_w)] = 0,\label{eq:tw frame 1}\\
    \frac{\diff ~}{\diff \eta}\left[AU(\shape U - c_w) + \frac{\Delta
    I}{\Fr^2}\right] = A - \tau.\label{eq:tw frame 2}
\end{gather}
Integrating the first of these equations gives $A(U - c_w) = Q_w$, where $Q_w$ is
the (constant) net flux in the frame of the wave. 
%Since both the mass $A$ and flux $AU$ are conserved quantities, $Q = 1-c$.
This allows us to substitute
\begin{equation}
    U = c_w + Q_w/A
    \label{eq:tw U}%
\end{equation}
into the second equation to eliminate $U$. 
After differentiating, rearranging and noting that $\mathrm{d}I/\mathrm{d}A =
D/\Delta$,
a single first-order ordinary
differential equation~(ODE) in terms of $A$ may be obtained:
\begin{equation}
%    \frac{\diff A}{\diff \eta} = \frac{A - \tau}{\Delta I'/\Fr^2 - \shape
%    Q_w^2/A^2 + c_w^2(\shape-1)},
%    \frac{\diff A}{\diff \eta} = \frac{A - \tau}{\Delta I'/\Fr^2 - Q_w^2/A^2 + \beta D -
%    \alpha Q_w/A}
    \frac{\diff A}{\diff \eta} = \frac{\tau(A) -
    A}{\lambda_w^-(A)\lambda_w^+(A)},
    \label{eq:tw ode}%
\end{equation}
where
\begin{equation}
    \lambda_w^\pm(A) = \frac{Q_w}{A} + \frac{\alpha}{2} \pm \frac{1}{\Fr}
    \sqrt{D + \Fr^2(\alpha^2/4 + \beta D)}
    \label{eq:lambda w}%
\end{equation}
are the system characteristics in the reference frame of the wave. In
both~\eqref{eq:tw ode} and~\eqref{eq:lambda w}, the closures $\tau$, $\alpha$
and $\beta$ are assumed
to have $U$ eliminated in favour of $A$, via~\eqref{eq:tw U}.

We seek a rising periodic solution to~\eqref{eq:tw ode} corresponding to a single roll
wave of some unknown amplitude and wavelength.
Continuous solutions to this ODE are necessarily monotonic, since the gradient
of $A$ in such circumstances can only change sign by passing through a fixed
point. Therefore, periodicity must be enforced at the peak of the wave by
connecting it with its tail via a
shock discontinuity. %, which we locate at $\eta = 0$, without loss of generality.
%By enforcing conservation of mass and momentum at the shock, the following jump
%conditions are obtained:
%%
%\begin{gather}
%    [A(U-c)]^+_- = 0,
%    \quad\mathrm{and}\quad \left[AU(\shape U - c) + \frac{\Delta
%    I(A)}{\Fr^2}\right]^+_-.
%\end{gather}
%%
%where $[f(A,U)]_-^+ \equiv \lim_{\eta\to 0^+}f(\eta) - \lim_{\eta\to 0^-}f(\eta) $.
The presence of a shock implies a change in the sign of one of the
characteristics at this transition.  For this to occur in a periodic solution,
the corresponding characteristic must reverse its sign change at some internal
point $(\eta, A, U) = (\eta_*, A_*, U_*)$, where the solution is continuous. 
Therefore, at this critical point, one of the characteristics are zero.
%, thereby causing the
%Jacobian of equations~\eqref{eq:tw frame 1}
%and~\eqref{eq:tw frame 2} to be singular, which ultimately leads to the
%vanishing of the denominator
%of the right-hand side of~\eqref{eq:tw ode}. 
This observation determines the wave speed.
By substituting 
\begin{equation}
    Q_w = A_*(U_*-c_w)
    \label{eq:Qw}%
\end{equation}
into~\eqref{eq:lambda w} and seeking the roots of the right-hand side, we obtain
two possible solutions
\begin{equation}
    %c = \shape \pm \frac{1}{\Fr}\sqrt{1 + \Fr^2\shape(\shape-1)}.
    %c_w = \shape U_* \pm \frac{1}{\Fr}\sqrt{\Delta I'(A_*) + \Fr^2\shape(\shape-1)U_*^2}.
    c_w = U_* + \frac{\alpha_*}{2} \pm \frac{1}{\Fr}\sqrt{D_* + \Fr^2\left(
    \frac{\alpha_*^2}{4} + D_* \beta_* \right)},
    \label{eq:cw}%
\end{equation}
labelling quantities evaluated at the critical point with `$*$' subscripts.  If
the negative branch of~\eqref{eq:cw} is taken, then $c_w < U_*$, implying that
$Q_w > 0$ from~\eqref{eq:Qw}.  From~\eqref{eq:lambda w}, $\lambda_w^+>0$ in this
case, so it must be $\lambda_w^-$ that changes sign at the critical point. Since
$\lambda_w^-$ is a decreasing function of $A$ for a (rising) roll wave solution,
this implies that $\lambda_w^- < 0$ upstream of the shock and $\lambda_w^- > 0$
downstream of it, rendering the shock inadmissible under the Lax entropy
condition.  The converse argument with signs reversed leads to an admissible
shock for the positive branch of~\eqref{eq:cw} with $\lambda_w^+$ changing sign
at the critical point.
%It is the positive branch of~\eqref{eq:cw} that must be taken to obtain roll
%wave solutions.

This leads to a singularity in~\eqref{eq:tw ode}, which must be
removed by ensuring
that the numerator of~\eqref{eq:tw ode} also vanishes.  This constrains the
critical point to lie on the curve 
\begin{equation}
    A_* = \tau(A_*,U_*),
    \label{eq:crit point drag condition}%
\end{equation}
which is exactly the
condition for steady uniform flow. An additional condition is needed to fix the
critical point for a particular flow. Roll waves that emerge from a flow fed by a
constant upstream source must deliver the same flux averaged over their
wavelength $\Lambda$ as the steady uniform base flow, measured in the laboratory
frame. Therefore,
\begin{equation}
    \frac{1}{\Lambda}\int_0^{\Lambda} AU \,\diff x = 1.
%    \quad\mathrm{or~equivalently,}\quad
%    \frac{1}{\lambda}\int_0^{\lambda} A \,\diff x = A_* + \frac{1-A_*U_*}{c_w},
    \label{eq:flux condition}%
\end{equation}
The amplitude and wavelength of each solution is contingent on the magnitude of
the shock, which must obey a set of jump conditions to ensure
conservation of mass and momentum over the discontinuity.
In particular, the condition for momentum is
\begin{equation}
    \left[ AU(\shape U - c_w) + \frac{\Delta I}{\Fr^2}\right]^+_- = 0,
    \label{eq:jump condition}%
\end{equation}
where $[f(\eta)]^+_-\equiv \lim_{\eta\to\eta^+} f(\eta)-\lim_{\eta\to\eta^-}f(\eta)$.

In both~\eqref{eq:flux condition} and~\eqref{eq:jump condition}, $U$ and
$c_w$
may be eliminated by using~\eqref{eq:tw U},~\eqref{eq:Qw} and~\eqref{eq:cw} to
give conditions on the flow area in terms of the critical point values.
Roll wave solutions for a given amplitude may then be constructed by numerically
integrating~\eqref{eq:tw ode}, subject to~\eqref{eq:Qw}
and~\eqref{eq:cw}, out from a critical point satisfying~\eqref{eq:crit point
drag condition}, in both directions, then using~\eqref{eq:jump condition} to
determine the upper and lower limiting points of the wave.
To satisfy the flux condition, this procedure may be automated and applied in
concert with a suitable root finding algorithm to isolate the critical point
that leads to the particular solution that obeys~\eqref{eq:flux condition}.
%Depending on the channel, functions for $\Delta$, $I$ and $D$ must be specified in terms
%of $A$, along with closures for $\tau$

This general method is employed in \S\ref{sec:numerics} to compute wave
solutions for debris flows in trapezoidal channels with the granular drag model
defined in~\eqref{eq:granular friction}
and~\eqref{eq:granular drag}.
Most of the geometric terms needed to close~\eqref{eq:tw ode}--\eqref{eq:jump
condition} in this case are given in \S\ref{sec:trapezoids}.
For completeness,
%In order to determine the range of admissible roll wave amplitudes, a form for
%$I$ must first be given and this is contingent on the channel geometry.
%Then Eq.~\eqref{eq:tw ode} can be integrated with respect to the limits set by
%this equation.
%
%For the trapezoidal case, we can find 
we provide formulae for the pressure integral terms here:
\begin{equation}
    I = \frac{wS_b s^2/2 + s^3/3}{wS_b + 1},\quad
    \frac{\mathrm{d}I}{\mathrm{d}A} 
    = \frac{A(wS_b + 1)}{\sqrt{4A(wS_b + 1) + (wS_b)^2}}
    = \frac{D}{\Delta}.
\end{equation}
After solving for $A$, the free surface elevation may be determined as
\begin{equation}
    s = \frac{1}{2}\left[
        \sqrt{(wS_b)^2 + 4A(wS_b + 1)} - wS_b
        \right].
\end{equation}

\end{document}